\documentclass[twocolumn]{aastex61}
\pdfoutput=1 
\usepackage{amsmath,amstext}
\usepackage[T1]{fontenc}
\usepackage{apjfonts} 
\usepackage[figure,figure*]{hypcap}
\usepackage{commath}

\shorttitle{Protostellar Hamburgers}
\shortauthors{Galv\'an-Madrid et al.}

\begin{document}

\title{On the Effects of Self-Obscuration in the (Sub-)Millimeter Spectral Indices \\ and Appearance of Protostellar Disks}

\correspondingauthor{Roberto Galv\'an-Madrid}
\email{r.galvan@irya.unam.mx}

\author[0000-0003-1480-4643]{Roberto Galv\'an-Madrid}
\affil{
Instituto de Radioastronom\'ia y Astrof\'isica (IRyA), UNAM, Apdo. Postal 72-3 (Xangari), Morelia, Michoac\'an 58089, Mexico. 
}

\author[0000-0003-2300-2626]{Hauyu Baobab Liu}
\affil{
European Southern Observatory (ESO), Karl-Schwarzschild-Str. 2, 85748, Garching, Germany.
}
\affil{
Academia Sinica Institute of Astronomy and Astrophysics, P.O. Box 23-141, Taipei 106, Taiwan.
}

\author[0000-0001-8446-3026]{Andr\'es F. Izquierdo}
\affil{
Instituto de F\'isica - FCEN, Universidad de Antioquia, Calle 70 No. 52-21, Medell\'in, Colombia.
}
\affil{
Instituto de Radioastronom\'ia y Astrof\'isica (IRyA), UNAM, Apdo. Postal 72-3 (Xangari), Morelia, Michoac\'an 58089, Mexico. 
}

\author[0000-0002-7997-2528]{Anna Miotello}
\affil{
European Southern Observatory (ESO), Karl-Schwarzschild-Str. 2, 85748, Garching, Germany.
}

\author{Bo Zhao}
\affil{
Max-Planck-Institut f\"ur extraterrestrische Physik (MPE), D-85748 Garching, Germany.
}

\author{Carlos Carrasco-Gonz\'alez}
\affil{
Instituto de Radioastronom\'ia y Astrof\'isica (IRyA), UNAM, Apdo. Postal 72-3 (Xangari), Morelia, Michoac\'an 58089, Mexico. 
}

\author[0000-0002-2260-7677]{Susana Lizano}
\affil{
Instituto de Radioastronom\'ia y Astrof\'isica (IRyA), UNAM, Apdo. Postal 72-3 (Xangari), Morelia, Michoac\'an 58089, Mexico. 
}

\author[0000-0003-2737-5681]{Luis F. Rodr\'iguez}
\affil{
Instituto de Radioastronom\'ia y Astrof\'isica (IRyA), UNAM, Apdo. Postal 72-3 (Xangari), Morelia, Michoac\'an 58089, Mexico. 
}

\begin{abstract}
In this paper we explore the effects of self-obscuration in protostellar disks with a radially decreasing temperature gradient and a colder midplane. We are motivated by recent reports of resolved dark lanes (`hamburgers') and  (sub)mm spectral indices systematically below the ISM value for optically thin dust $\alpha_{\rm ISM} =3.7$.
We explore several model grids, scaling disk mass and varying inclination angle $i$ and observing frequency $\nu$ from the VLA Ka band ($\sim 37$ GHz) to ALMA Band 8 ($\sim 405$ GHz). We also consider the effects of decreasing the index of the (sub-)mm dust opacity power law $\beta$ from 1.7 to 1. 
We find that a distribution of disk masses in the range $M_{\rm disk} = 0.01-2~M_\odot$ is needed to reproduce the observed distribution of spectral indices, and that assuming a fixed $\beta =1.7$ gives better results than $\beta=1$. 
A wide distribution of disk masses is also needed to produce some cases with $\alpha <2$, as reported for some sources in the literature. Such extremely low spectral indices arise naturally when the selected observing frequencies sample the appropriate change in the temperature structure of the optically thick model disk. 
Our results show that protostellar disk masses could often be underestimated by $> \times10$, and are consistent with recent hydrodynamical simulations.  
Although we do not rule out the possibility of some grain growth occurring within the short protostellar timescales, we conclude that self-obscuration needs to be taken into account.
\end{abstract}


\section{Introduction} \label{sec:intro}
Evidence has accumulated showing that disk formation occurs during the protostellar stages: the so-called class 0 and I Young Stellar Objects \citep{Rodriguez98,Jorgensen09,Choi2010,Lee10,Murillo13,Sakai14}. For a review, see \cite{Li14}. The mass reservoir in these disks is one of the most important initial conditions for the posterior  `protoplanetary disks', or class II YSOs. It is possible that protostellar disks build up in mass through accretion from their envelope until they reach their class II mass. However, inferred gas masses in protoplanetary disks vary over a wide range, from $\sim 10^{-5}$ to $10^{-1}~M_\odot$ \citep{Bergin13,Pascucci16,McClure16,Ansdell16,Miotello17}.  

A number of detailed hydrodynamical simulations progressively including more physics show that protostellar disks have a variety of properties, and that often they are relatively massive compared to their parent envelope and the corresponding stellar mass, reaching masses up to a few $\times 0.1~M_\odot$ for an envelope of one solar mass \citep{Machida14,Vorobyov15,Ilee17,Zhao18,Bate18}. 

Large protostellar disk masses may seem inconsistent with previous observational determinations of $M_{\rm disk} \sim 0.01$ to $0.1~M_\odot$ \citep[e.g.,][]{Jorgensen09,Tobin13}, but mass estimations are sensitive to the selection of slope $\beta$ and normalization $\kappa_0$ of the (sub)mm dust opacity power law $\kappa_\nu = \kappa_0 \nu^\beta$, and could often be underestimated by $\times 10$ or more \citep{Evans17}. Recent observations have indeed shown the existence of massive, self-gravitating protostellar disks \citep{Tobin16}. 
Further evidence for a systematic underestimation of protostellar disk masses could come from observations of (sub)mm spectral indices significantly below the optically-thin ISM value $\alpha_{\rm ISM} = 2 + \beta_{\rm ISM} = 3.7$. To date, several protostellar objects have been reported to have inferred $\beta \approx 0$ to 1.5 \citep{Chiang2012,Miotello14,LiLiu17,Gerin17,Sheehan2017}. These low spectral indices are often interpreted as a signature of significant dust grain growth \citep{Miotello14}, or a combination of some grain growth and large optical depths \citep{Gerin17}. However, it is not clear if the few $\times 10^5$ yr lifetime of protostars is long enough for dust to evolve significantly \citep{Ormel09,Wong16}.
Due to the potentially very efficient inward migration of grown dust, it is also not clear whether or not observations can be sensitive to grain growth signatures during the embedded phase of protostars \citep{Vorobyov18}.

An alternative interpretation for the observed low spectral indices is that there is significant self-obscuration due to protostellar disks being partially optically thick, particularly at higher frequencies, i.e., that they have masses $M_{\rm disk} >> 0.01~M_\odot$. This effect is enhanced if there is a temperature gradient. \cite{LiLiu17} reported SMA observations where the envelope is subtracted from an unresolved disk component in the visibility domain, and showed that a simple two-temperature model with a hot, relatively massive region inside a colder, also massive one was enough to reproduce the low (sub)mm spectral indices in their disk sample. 
In their interpretation, the hot, inner component -- the inner disk -- is hidden at all frequencies $> 100$ GHz, and only starts contributing to the observed flux at lower frequencies, thus flattening the  spectral index. More recently, \cite{Lee17b} presented astonishing ALMA images of an edge-on protostellar disk with 8 au resolution which show a prominent dark lane. This characteristic `hamburger' morphology proves the existence of self-obscuration at (sub)mm wavelengths in this particular object. It also suggests the existence of a disk that is warmer at inner radii and colder in its midplane. A similar morphology is also evident, although not well resolved, in the images of the `Butterfly Star' presented by \cite{Graefe13}. 

In this paper, we further explore the hypothesis that 
the systematically low spectral indices observed in protostellar disks are mainly caused by self-obscuration. In Section \ref{sec:model} we explain our fiducial physical model as well as the radiative transfer calculations that were performed. In Section \ref{sec:grid_of_models} we describe the grids of models that were built to explore parameter space. Section \ref{sec:trends_alpha} reports the trends in the model spectral indices. Section \ref{sec:comparison} compares the ensemble of model spectral indices to the observational sample of \cite{LiLiu17}. Section \ref{sec:discussion} presents a discussion of the implications of our results. In section \ref{sec:conclusions} we give our conclusions. 

\section{Fiducial Model} \label{sec:model}

\subsection{Model Structure} \label{sec:model_structure}

\begin{figure*}[!ht]
\begin{center}
\includegraphics[width=0.33\textwidth]{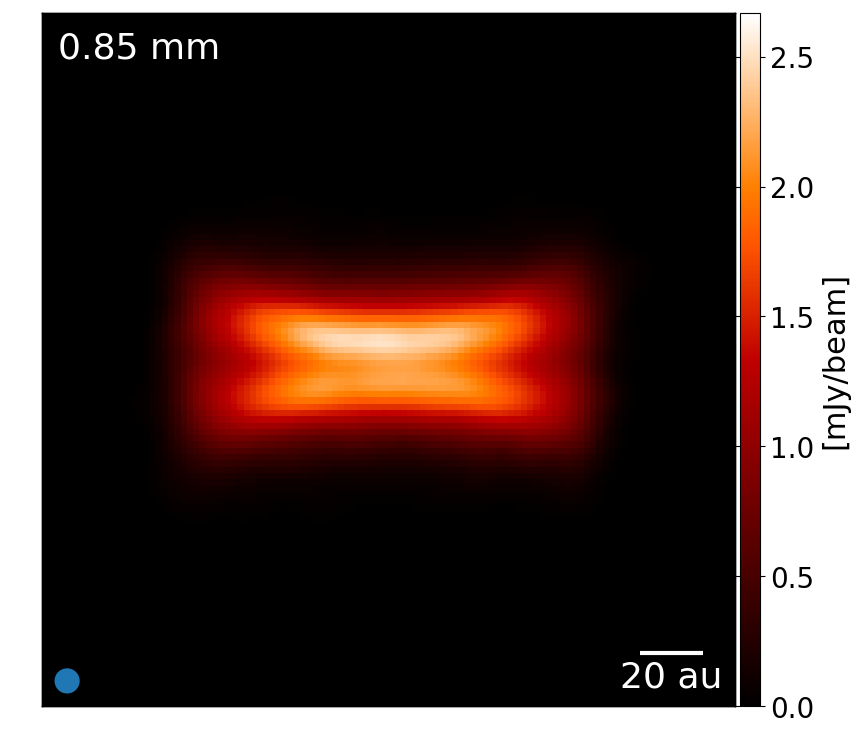}
\includegraphics[width=0.33\textwidth]{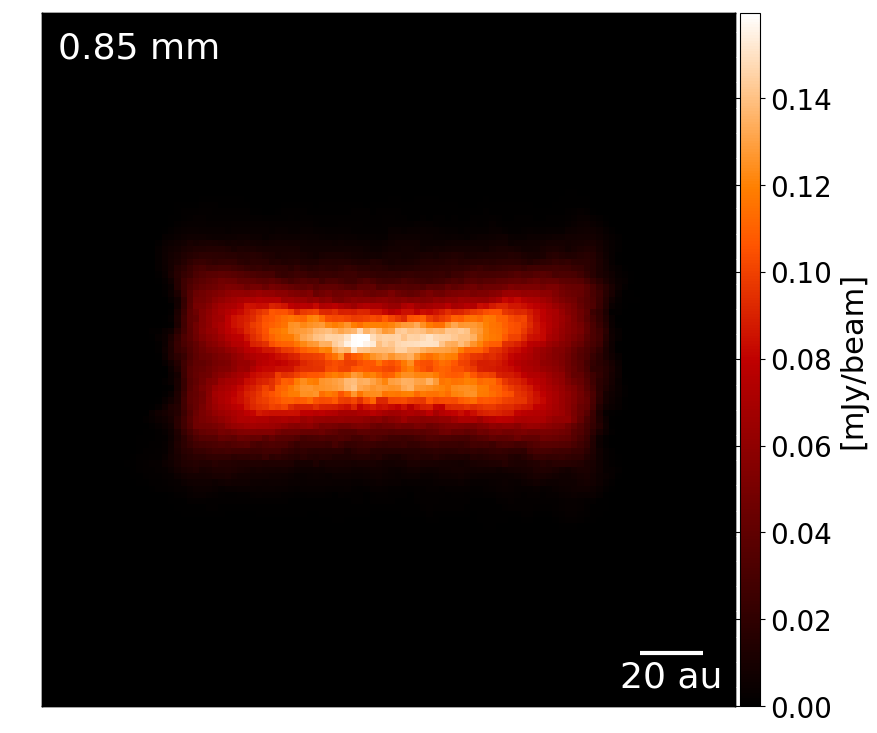}
\includegraphics[width=0.33\textwidth]{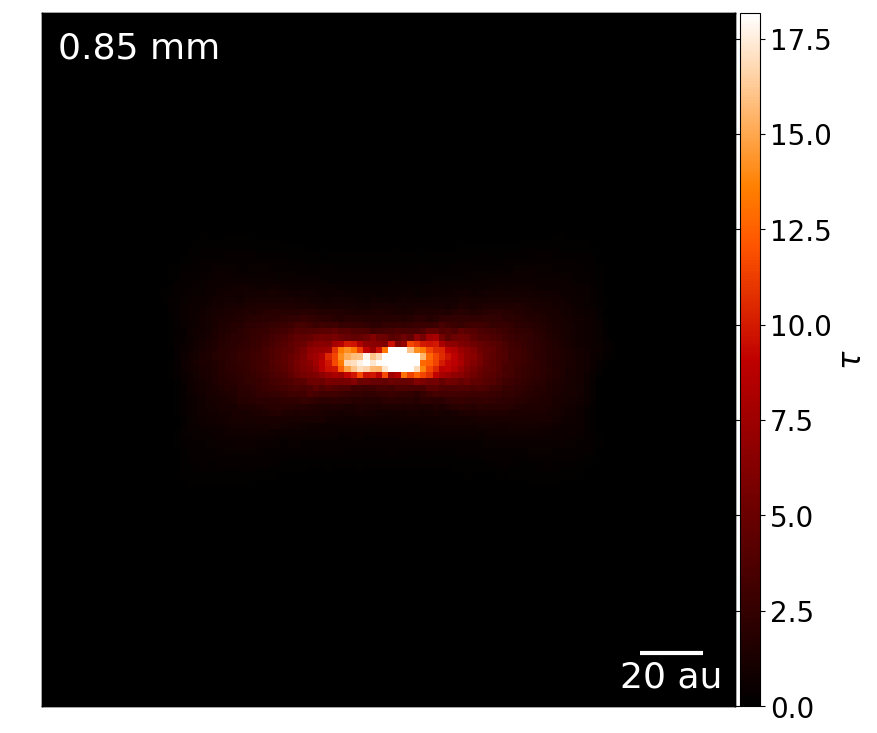}
\caption{Fiducial model that approximately reproduces the resolved `hamburger' reported by \cite{Lee17b}. {\it Left:} intensity map at 0.85 mm (352.7 Ghz) convolved with a 20 mas FWHM gaussian beam. {\it Center:} same as {\it left} but unconvolved. {\it Right:} corresponding optical depth $\tau$ map. \label{fig:benchmark}}
\end{center}
\end{figure*}

Given the current uncertainties on the detailed structure of disks in class 0/I YSOs, we followed an empirical approach to find a fiducial model that then could be scaled in mass and adjusted in inclination angle. This fiducial model was built to approximately match the recently reported observation of a resolved dark lane in the almost edge-on disk of the class 0 YSO IRAS 05413--0104 \citep{Lee17b,Lee18}, which drives the famous HH 212 jet \citep{Lee17a}. This `hamburger' is the first to be fully resolved at (sub)mm wavelengths, thanks to the power of ALMA long-baseline observations with a resolution of $\sim 20$ mas (8 au at a distance of 400 pc). At these resolutions, (sub)mm continuum images trace the `true' disk without significant contributions from the envelope \citep{Lee17b}.  

We use a standard accretion disk \citep{Pringle81,Dalessio98,Keto10} with a density $\rho$ that goes as:
\begin{subequations}
\begin{gather}
\begin{flalign}
& \rho = \rho_R \rho_Z, \\
& \rho_R = \rho_0 (R_d/R)^{2.25}, \\
& \rho_z = \exp (-z^2/2H^2).
\end{flalign}
\end{gather}
\end{subequations}
where $\rho_0$ and $R_d$ are normalization constants, $R$ and $z$ are cylindrical coordinates, and $H(R)\propto R^{1.25}$ is the scale height of the flared disk \citep{Whitney03,Keto10}. 
The models presented in this paper are truncated at a radius $R=68$ au.
 
For the temperature profile we adopt the following prescription \citep{Pringle81,Keto10}:
\begin{subequations}
\begin{gather}
\begin{flalign}
& T = T_R T_z,  \\
& T_R = [T_0 R^{-3}(1-(R_\star/R)^{1/2})]^{1/4},   \\
& T_z = \exp [-(\abs{z}-H)^2/2H^2].
\end{flalign}
\end{gather}
\end{subequations}
where $T_0$ is a normalization constant and $R_\star$ is the stellar radius, which we set to $\times 10$ the main sequence value for a $0.3~M_\odot$ star, thus $R_\star = 3.8~R_\odot$, given that accreting protostars should be significantly bloated compared to ZAMS stars of the same mass due to their high accretion rates \citep{HosokawaOmukai10}. 

Figure \ref{fig:model_cuts} in Appendix \ref{sec:model_structure} shows the central density and temperature cuts for our benchmark across the disk midplane and perpendicular to it. 
We explicitly include a colder midplane, as indicated by recent observational determinations in more evolved objects \citep[e.g.,][]{Guilloteau16}. This is expected if protostellar irradiation, including accretion heating, is dominant. The temperature decrease for $\abs{z} > H$ was set as a simple way of having a smooth, non-divergent temperature distribution. The exact shape of $T(z)$ for $\abs{z} > H$ has very little effect on the emissivity due to the very low densities in these regions. 
We set a minimum temperature in the model grid $T_{\rm min}=10$ K, and verified that the results are not affected by changing this value between 5 and 20 K. The profile normalizations are such that temperature and density of H$_2$ particles are 121 K and $8.2 \times 10^{11}$ cm$^{-3}$ at 10 au in the midplane, respectively. Total gas masses are calculated assuming an H$_2$ gas-to-dust mass ratio of 100 and $M_{\rm gas} = 1.4M_{H_2}$. 

\subsection{Radiative Transfer} \label{sec:model_rt}

The dust continuum radiative transfer was calculated using LIME \citep{Brinch10}.
For a given line of sight, the emergent monochromatic intensity is: \\
\begin{equation}
I_\nu(\tau_\nu) = I_\nu(0) \exp(-\tau_\nu) + \int_0^{\tau_\nu}\exp(-\tau_\nu + \tau_\nu^\prime) B_\nu(\tau_\nu^\prime) d\tau_\nu^\prime,
\end{equation} \\
where $I_\nu(0)$ is the background, taken as the CMB intensity, the local source function is equal to the Planck function with temperature $T$ at $\tau_\nu^\prime$, and the optical depth is $\tau_\nu = \int \kappa_\nu \rho dl$. Please note that the observed monochromatic intensity is proportional to the mass per unit area $\Sigma = \int \rho dl$ only when the emission is optically thin ($\tau_\nu <<1$) and $T$ and $\kappa_\nu$ do not vary along the line of sight. In this often assumed limiting case, the estimated dust mass is $M \propto \kappa^{-1}$, thus an uncertainty in the opacity $\sigma_\kappa$ translates into a mass uncertainty $\sigma_M = (M/\kappa)\sigma_\kappa$. 
In the case of temperature variations or large optical depths in the line of sight this simple case is not valid anymore. For the calculations in this paper, 
a dust opacity power law $\kappa_\nu \propto \nu^\beta$, $\beta = 1.7$, and $\kappa_{1300\mu {\rm m}} = 1$ cm$^2$ g$^{-1}$ are used throughout, i.e., we assume typical ISM dust \citep{Draine06}. 
In Section \ref{sec:comparison} we also consider the case with $\beta = 1$. 

The physical models are created and plugged into LIME using the tools presented by \cite{Izquierdo18}.  
There are two grids to consider: the `native' model grid and the LIME grid. The former grid is cartesian and has a length of 200 au divided in 200 cells per dimension. In contrast, LIME's grid for the ray-tracing is quasi-random with a larger density of points in regions of larger mass density.  The assignment of densities and temperatures from the native grid to the LIME grid is also done by the `grid ingestor' presented and tested in \cite{Izquierdo18}. For our current application, we found that 5000 LIME points are enough to produce correct results. Small  fluctuations in the output images at the level of a few \% (see Fig. \ref{fig:benchmark}) are corrected by calculating each of them ten times and taking their median. 

To avoid the allocation of too many LIME points in the innermost regions with a size of the order of the native-grid cell size, a minimum scale has to be defined in LIME. We set it to 1 au. The missing flux at these scales is negligible except for the lower-mass, lower frequency models, where it amounts to $\sim 1-5~\%$ of the total. The output images are resampled with a pixel size of 5 mas, or 2 au at 400 pc.

A small background intensity $\sim 10^{-8}$ to $10^{-7}$ Jy/pix in the model images arises from the small background density left in the model grid outside the disk. Therefore, this background intensity only depends on frequency and it was exactly subtracted from the model images when fluxes were calculated. In real interferometric images this background would not appear since it would naturally be filtered out.

\begin{deluxetable}{c|c}[ht!]
\tablecaption{Modelled Bands\tablenotemark{a} \label{tab:table1}}
\tablehead{
\colhead{Band} & \colhead{Frequency} \\
\colhead{VLA/ALMA} & \colhead{[GHz]}
}
\startdata
Ka & 37 \\
Q & 43  \\
Band 3 & 98 \\
Band 4 & 145  \\
Band 5 & 203  \\
Band 6 & 230  \\
Band 7 & 340  \\
Band 8 &  405 \\
\enddata
\tablenotetext{a}{We set the frequencies to those used in the class 0/I sample of \cite{LiLiu17}. We also included other standard ALMA bands. For the latter case the frequency was selected as the default continuum setup from the ALMA technical handbook (\url{https://almascience.eso.org/}).}
\end{deluxetable}

\subsection{Benchmark} \label{sec:benchmark}

The total mass in our fiducial model is $M_{\rm disk} = 0.15~M_\odot$. With an inclination angle $i=87^\circ$\footnote{We define the inclination angle such that the disk is viewed edge-on at $i=90^\circ$ and face-on at $i=0^\circ$} we were able to reproduce the total flux ($\approx 140$ mJy at 850 $\mu$m), width of the dark lane ($\approx 40$ mas), and average brightness contrast of the upper and lower intensity peaks ($\approx 1.14$) of the resolved `hamburger'.  Figure \ref{fig:benchmark} shows the corresponding model image and optical depth $\tau_{\rm 850\mu m}$ map.  

We emphasize that it is not our purpose to reproduce the observations of \cite{Lee17b} to a fine-detail level, but that creating such model served as a benchmark test to the presented study of the systematic dependence of the (sub)mm spectral indices on the disk mass and inclination angle. Also, we intentionally did not pass the models through an assumption of the interferometric instrumental response, since future surveys of protostellar disks could have angular resolutions ranging from $\sim 10$ to 1000 mas and dissimilar noise levels. 

We performed tests of the effects of independently increasing the disk radius and adding an exponential tapering in the density profile of the outer disk, since such tapering helps to better reproduce the outer-disk appearance \citep[e.g.,][]{Lee17b}.
Increasing the disk truncation radius by $20~\%$ increases the Q-to-B6 and B3-to-B7 angle-integrated spectral indices in the benchmark model by only $\sim 4 ~\%$. On the other hand, an exponential tapering in the outer disk increases the local density at large radii. 
Figure \ref{fig:tapered_model} in Appendix \ref{sec:fiducial_structure} shows a density cut and the resulting 0.85 mm image of such a tapered model. The flux in the latter is only $1.5\%$ higher than in the untapered model.
The effect on the spectral indices is a small decrement of $1~\%$ to $3~\%$, depending on frequency.

\section{Grids of Models}  \label{sec:grid_of_models}

With the fiducial model defined, we created several grids of model continuum images by varying the model density normalization (its mass), observer's inclination angle, and frequency. A first illustrative grid, whose results are described in Section \ref{sec:trends_alpha}, was created with 
disk masses varied as $M_{\rm disk} = 0.05$, 0.15, 0.25, and 0.35 $M_\odot$. 

In all grids, the inclination angle from the line of sight $i$ was varied from 90 to $0^\circ$ in steps of $3^\circ$. This fine angle sampling was needed for the comparison to observational results presented in Section \ref{sec:comparison}. Also, the l.o.s optical depth $\tau$ varies more rapidly close to $i \sim 90^\circ$. The selected angle interval is enough to capture the changes of spectral index $\alpha$ with angle even when the system is close to edge-on. 

Frequencies were selected from the observational sample of \cite{LiLiu17}, which correspond to standard `continuum' settings at the VLA and SMA\footnote{The Submillimeter Array is a
joint project between the Smithsonian Astrophysical Observatory
and the Academia Sinica Institute of Astronomy and
Astrophysics and is funded by the Smithsonian Institution and
the Academia Sinica \citep{Ho04}.} interferometers. We also calculated model images at standard ALMA bands that could be used to systematically observe class 0 and I protostars in the near future. Table \ref{tab:table1} summarizes the used frequencies and their band naming convention. 

The statistical comparison to observations presented in Section \ref{sec:comparison} required the creation of a second grid of models with a more ample mass range. A  third grid of models with the same masses as the second but with $\beta = 1$ is also considered. In total, our grids consist of 16616 model images. 

\section{Results} \label{sec:results}

\subsection{Trends in Appearance and Spectral Indices} \label{sec:trends_alpha}

\begin{figure*}[!ht]
\gridline{
\fig{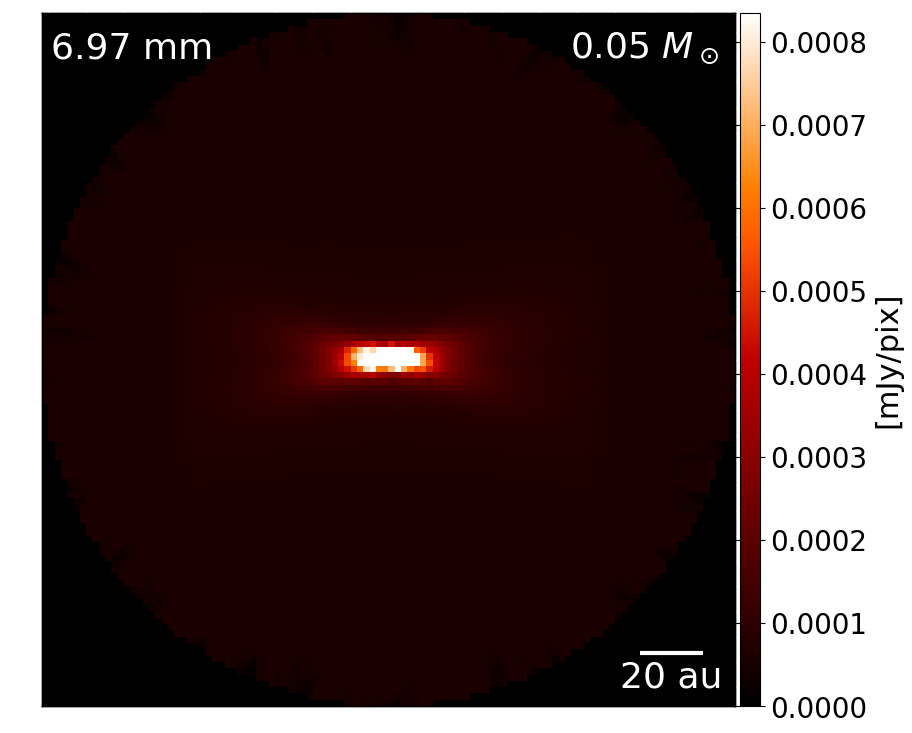}{0.32\textwidth}{}
\fig{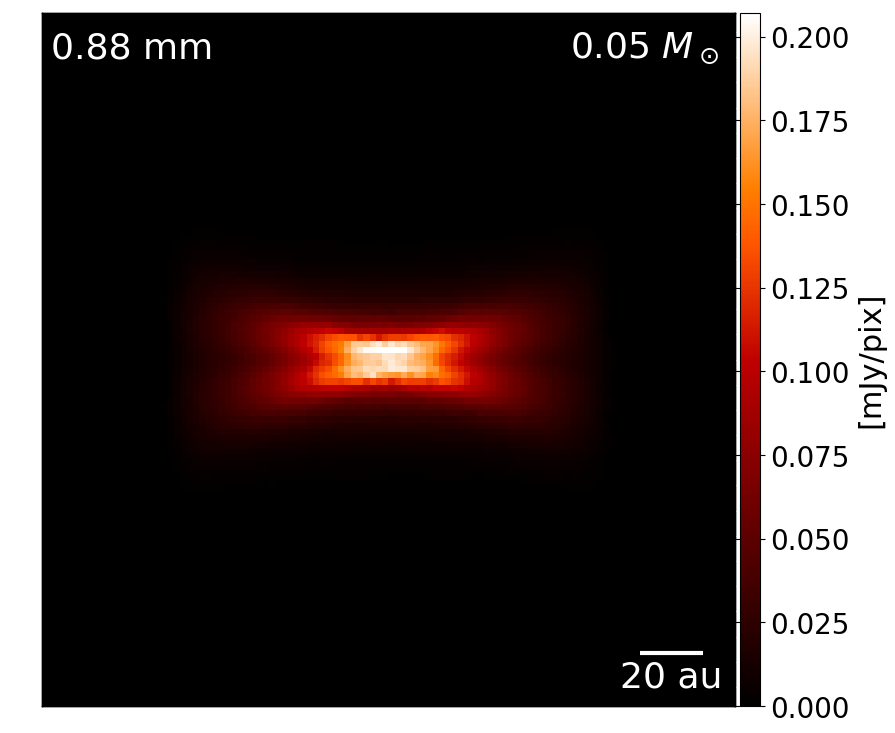}{0.32\textwidth}{}
\fig{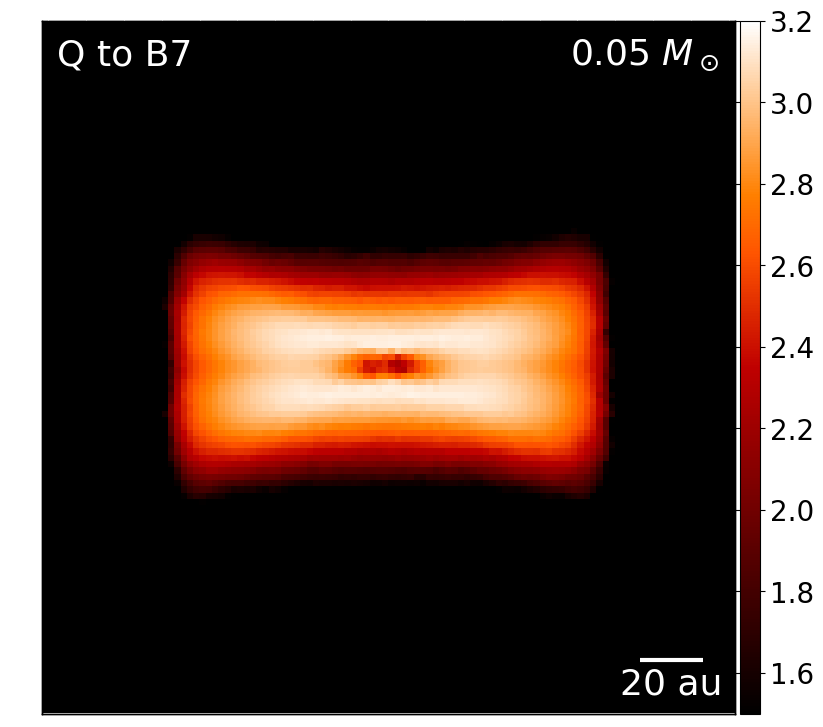}{0.30\textwidth}{}
          }
\gridline{
\fig{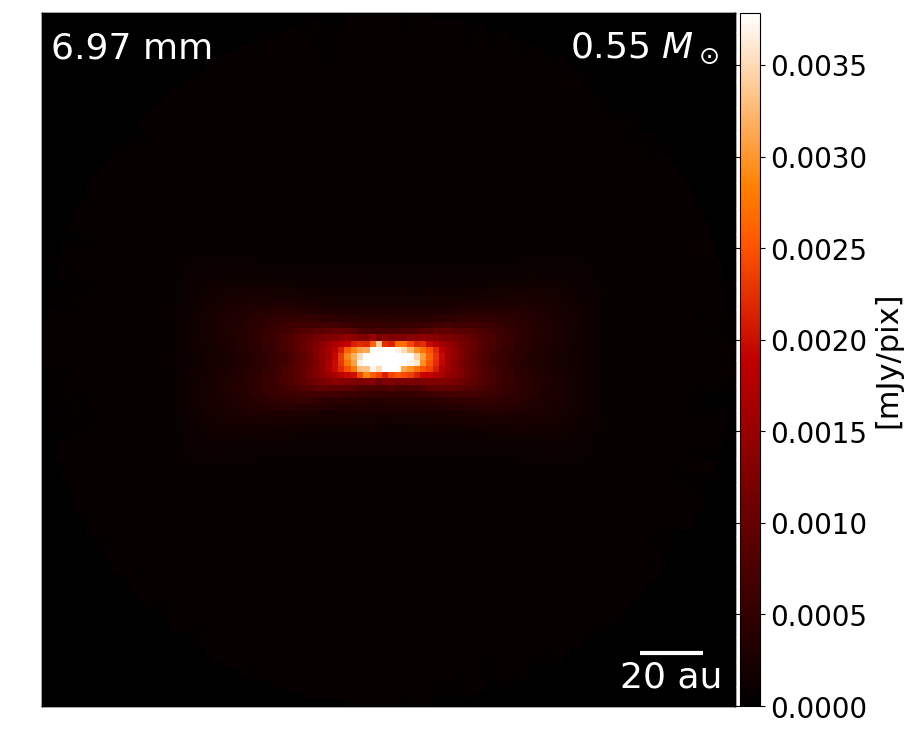}{0.32\textwidth}{}
\fig{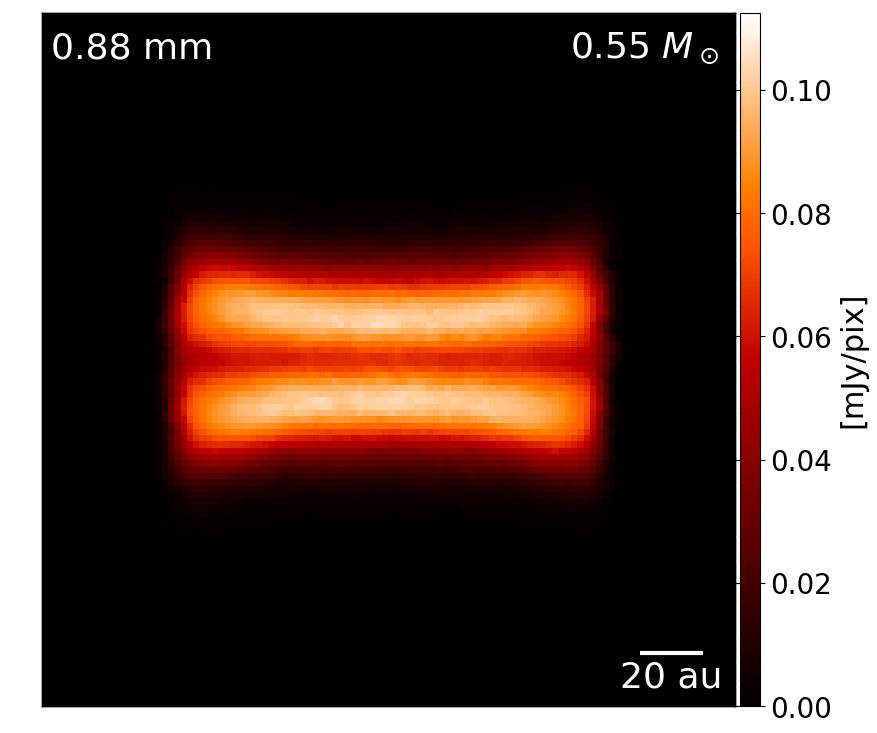}{0.32\textwidth}{}
\fig{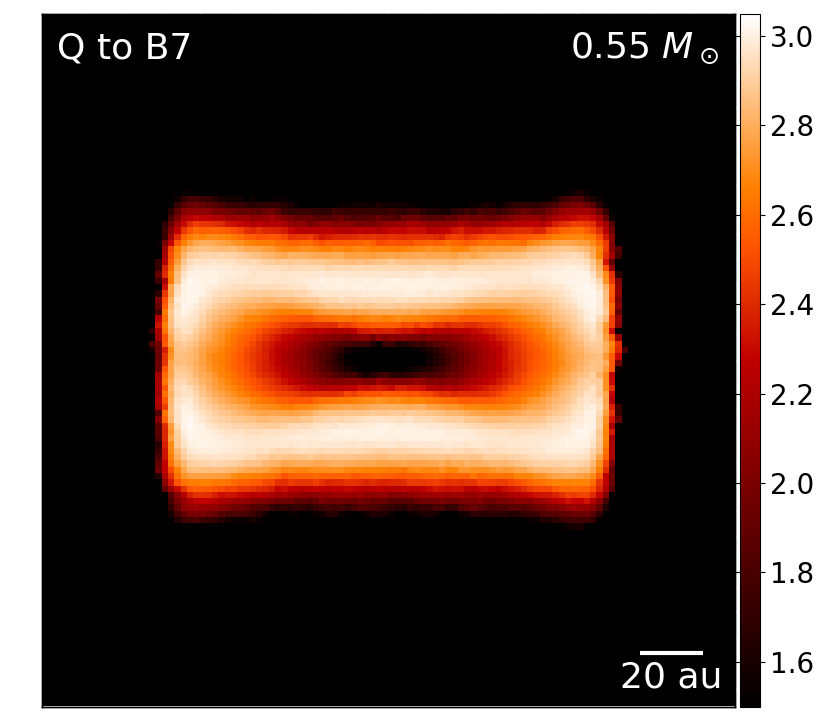}{0.30\textwidth}{}
          }
\caption{
Selected model and spectral index images. {\it Top row:} model with $M_{\rm disk}=0.05~M_\odot$. {\it Left:} Q-band image. {\it Center:} Band 7. {\it Right:} corresponding spectral index map. 
{\it Bottom row:} same as top row but for $M_{\rm disk}=0.55~M_\odot$.
\label{fig:selected_images}}
\end{figure*}

The following description of results correspond to the first grid of models defined in the previous Section. 

Our results confirm the intuitive expectation that prominent dark lanes due to self-absorption within a temperature structure such as the one we use are preferably seen when one or more of the following conditions apply: large masses, high frequencies, and large inclination angles. All of these conditions help in producing large continuum optical depths $\tau$ over increasingly larger fractions of the image (see Figs. \ref{fig:benchmark} and \ref{fig:selected_images}). 

At the lowest frequencies that we examine (the VLA Ka band, $\sim 37$ GHz) a prominent dark lane is not developed even in the most massive $M_{\rm disk} = 0.35~M_\odot$, edge-on models. In contrast, at the highest frequencies (ALMA Band 8, 405 GHz), even the lowest mass $M_{\rm disk} = 0.05~M_\odot$ edge-on model presents indications of a dark lane, although not prominently. 
In the most used ALMA bands 6 and 7, a prominent dark lane as wide as $\sim 50$ mas is seen in the $M_{\rm disk} = 0.35~M_\odot$ edge-on images, but there is just a hint of it in the  $M_{\rm disk} = 0.05~M_\odot$ ones. The currently lowest-frequency ALMA band (B3, 98 GHz) shows the signature dark lane only for disk masses $> 0.15~M_\odot$, and a lack of it otherwise. 

That a prominent dark lane is not present in fully resolved observations does not mean that there is not {\it some} self-obscuration. Interestingly, the first effects of self-obscuration can be more readily diagnosed from 
the spectral indices $\alpha = \log (F_2 / F_1) / \log (\nu_2 / \nu_1)$ obtained from angle-integrated intensities (i.e., fluxes).
The reason for this is that 
as the mass and thus the optical depth are increased, then the intensity in a given line of sight grows more rapidly for the lower- than for the higher frequency. This decreases $\alpha$ in the given line of sight, and can be measurable in the angle-integrated $\alpha$ if observational errors are small enough. 
As self-obscuration builds up by increasing mass, frequency, and/or l.o.s angle, a respectively larger fraction of the image suffers the previously described process until it becomes very optically thick (see Fig. \ref{fig:selected_images}). The resolved self-obscuration appears first as a dark lane toward the center of the disk midplane, and extends radially and vertically as the effect becomes more significant. In reality, the detailed geometry of how this happens could be different if our simple model assumption does not hold, but it is reassuring that it it is consistent with the first fully resolved image presented by \cite{Lee17b}. Low spectral indices can also be produced when dust emission in one or more of the observed bands is out of the Rayleigh-Jeans limit ($h\nu << k_{\rm B}T$). This effect can be appreciated toward the outskirts of the model in the right panels of Fig. \ref{fig:selected_images}, but it contributes very little to the spectral index of the entire object. However, its consideration may become relevant as fully resolved images become more common. 

\begin{figure*}[!ht]
\begin{center}
\includegraphics[width=0.49\textwidth]{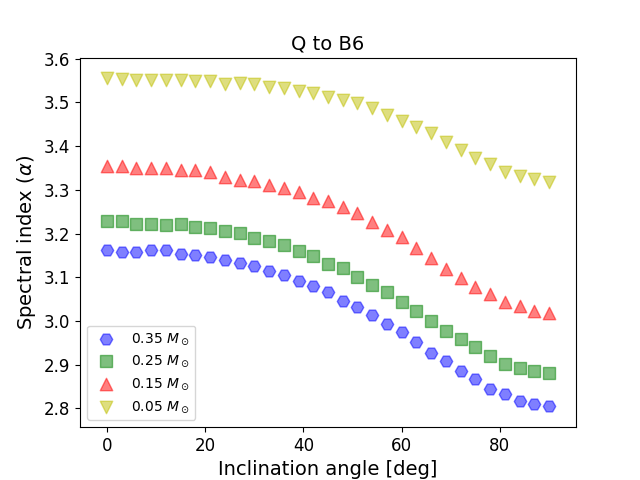}
\includegraphics[width=0.49\textwidth]{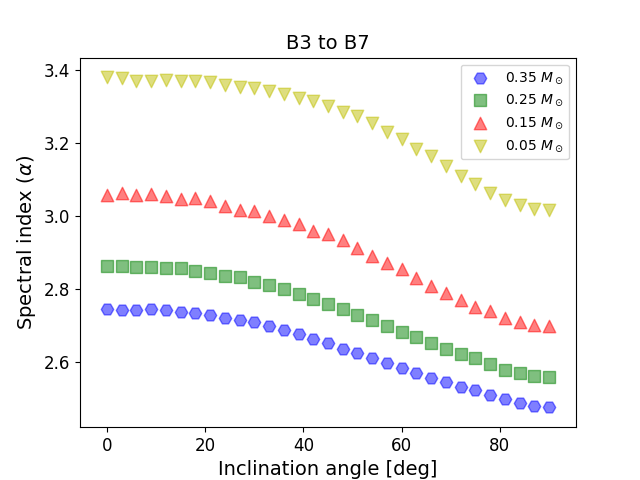}
\caption{Spectral indices as a function of inclination angle for representative, often-used combinations of bands. Different colored symbols are for different model masses.  \label{fig:representative_spix}}
\end{center}
\end{figure*}

Figure \ref{fig:compilation_alpha} in the Appendix \ref{sec:compilation} shows a compilation of the spectral indices for pairs of bands that are consecutive in frequency. Even though this would not be the best thing to do observationally because errors in the spectral indices would be larger than for bands that are more separated in frequency, we present the data in this way so the increasing importance of the effect with increasing frequency can be fully appreciated. Errors in the pure model spectral indices are mainly determined by the small leftover fluctuations in the median images and the 1 au radius central hole in the LIME calculations (see Section \ref{sec:model_rt}). They are $\sigma_{\alpha, \mathrm{model}} \sim 0.02$ for the lower-frequency, lower-mass calculations, and rapidly decrease to negligible levels for higher masses and frequencies. 

Figure \ref{fig:representative_spix} shows the spectral indices for two pairs of bands that are among the most commonly used in the literature, mostly in studies of more evolved protoplanetary disks \citep[e.g.,][]{CG16,Tazzari16}: VLA Q to ALMA Band 6 $\alpha_{\rm Q-B6}$ and Band 3 to Band 7 $\alpha_{\rm B3-B7}$. 
For the lower-mass $M_{\rm disk} = 0.05~M_\odot$ model, $\alpha_{\rm Q-B6}$ varies from 3.56 to 3.32 when $i$ is varied from $0^\circ$ to $90^\circ$. The respective variation for the most massive disk $M_{\rm disk} = 0.35~M_\odot$ is $\alpha_{\rm Q-B6} = 3.16$ to 2.81. On the other hand, $\alpha_{\rm B3-B7}$ varies with angle from 
3.38 to 3.02 for the least massive disk, and from 2.75 to 2.48 for the most massive disk. 

Assuming a flux calibration error of $10\%$ at each band, typical observational uncertainties in these two spectral indices would be $\Delta \alpha \sim 0.1$. 
Therefore, angle-integrated measurements of $\alpha_{\rm Q-B6}$ at angles closer to edge-on than face-on would be distinguishable from the value for optically-thin ISM dust $\alpha_{\rm ISM} = 3.7$ for all disk masses $> 0.1~M_\odot$. 
Similarly, observations of systems close to face-on would only have a  $\alpha_{\rm Q-B6}$ clearly discernible from the ISM value for the most massive disks. The effects of self-obscuration are more readily  noticeable for $\alpha_{\rm B3-B7}$, which is clearly below $\alpha_{\rm ISM}$ at high significance for all masses and inclination angles, except possibly for the least massive, face-on model.  

From the above described trends, it can be concluded that spectral indices between all ALMA bands are significantly affected by self-obscuration for all models except the least massive one, regardless of inclination. To allow the reader for more specific comparisons of their interest, a table with fluxes is given in the online-only version of this paper. 

\subsection{Comparison to Observational Sample} \label{sec:comparison}

\subsubsection{(Sub)millimeter spectral indices from model grids} \label{sec:gridalpha}

In this section, we further investigate the effect of viewing angles and the selection of observational frequency bands on the distribution of spectral indices. 
To do so, we linearly interpolated the model grid originally sampled every $3^\circ$ to 180 different inclination angles $i$ which uniformly sample $\cos i$ in the range [0,1]. 

\begin{figure*}[!t]
   \begin{tabular}{c}
     \vspace{-0.5cm}\includegraphics[width=\textwidth]{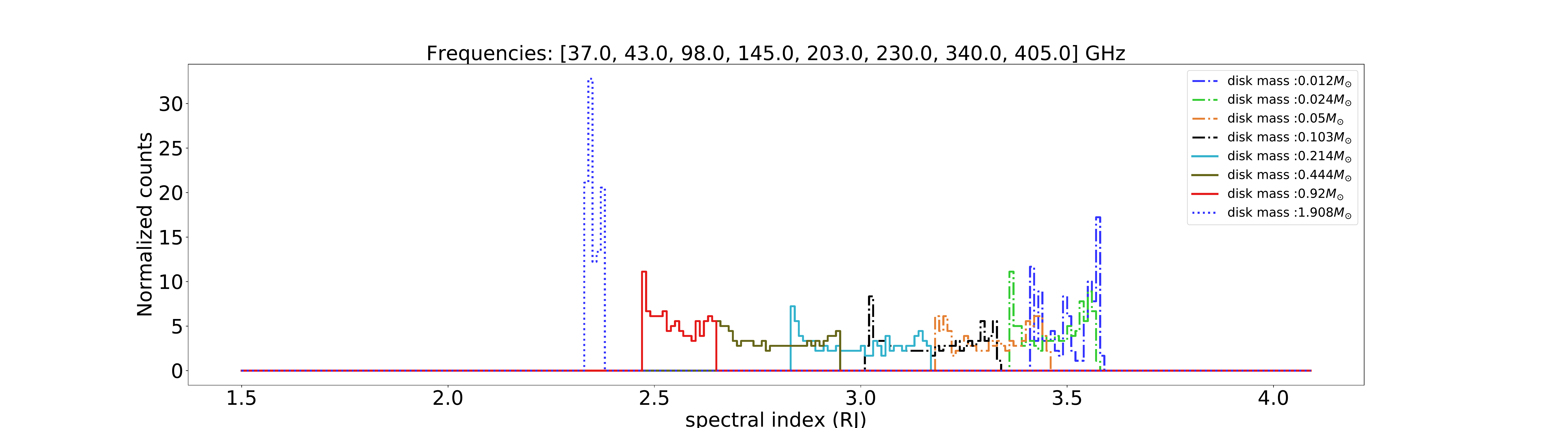}\\
     \vspace{-0.5cm}\includegraphics[width=\textwidth]{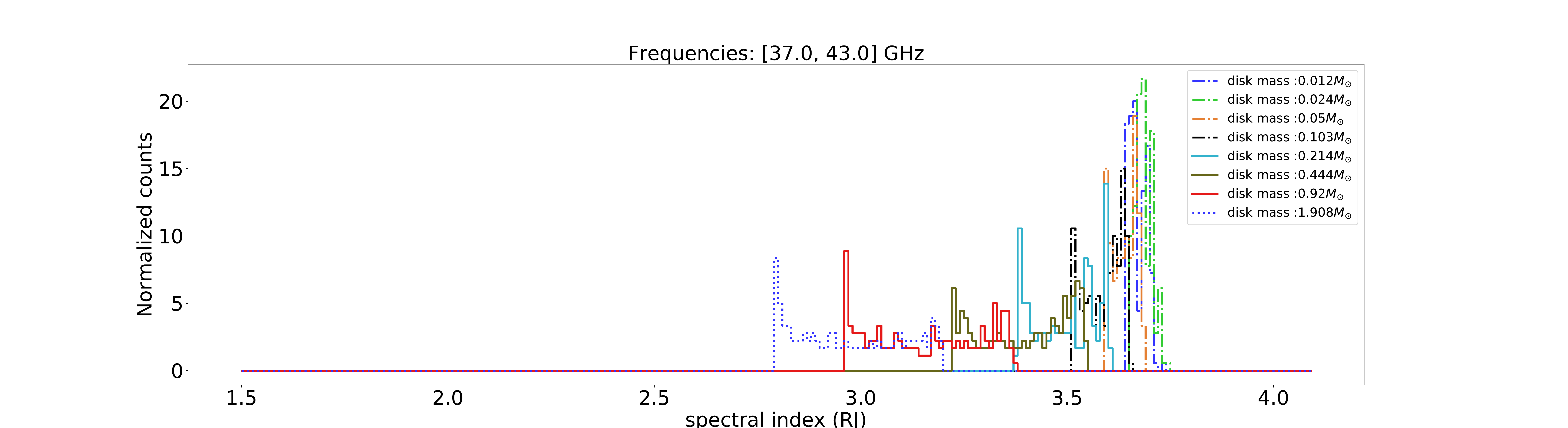}\\
     \vspace{-0.5cm}\includegraphics[width=\textwidth]{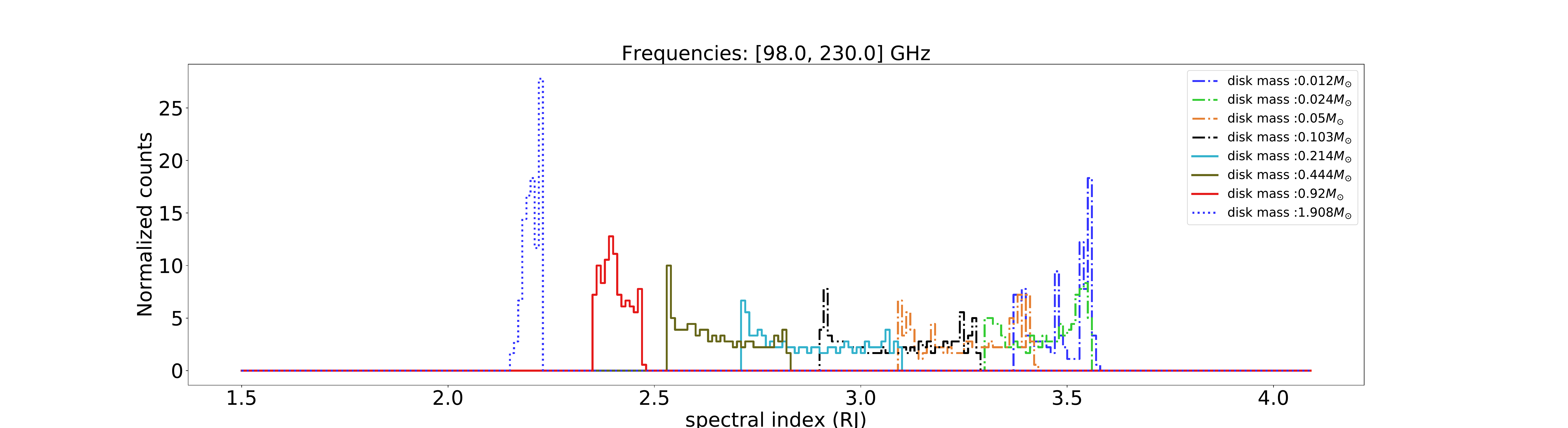}\\
     \vspace{-0.5cm}\includegraphics[width=\textwidth]{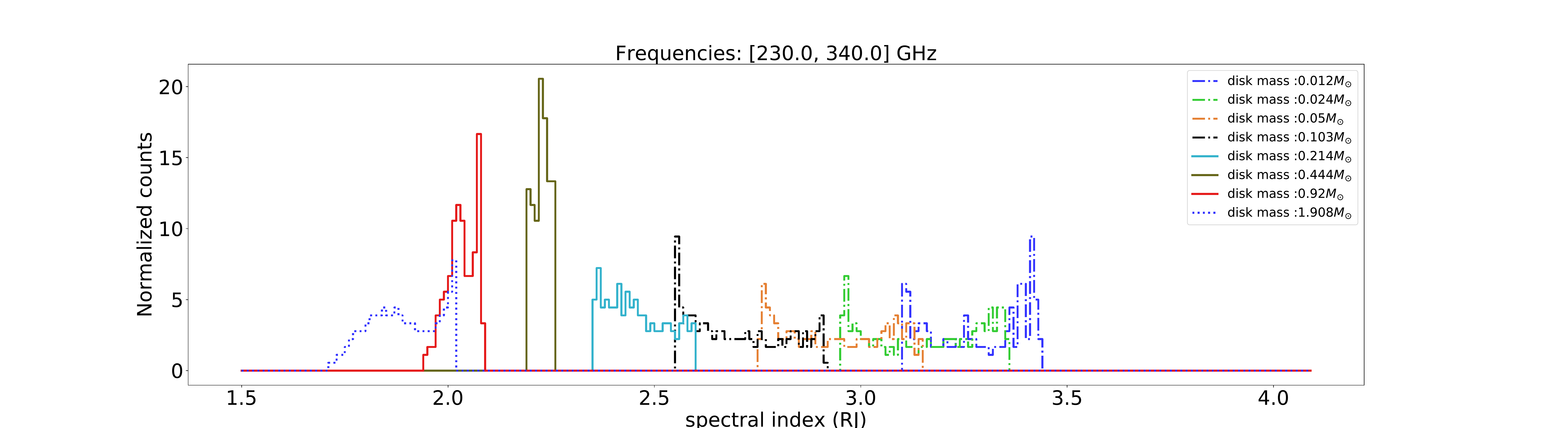}\\
     \vspace{-0.3cm}\includegraphics[width=\textwidth]{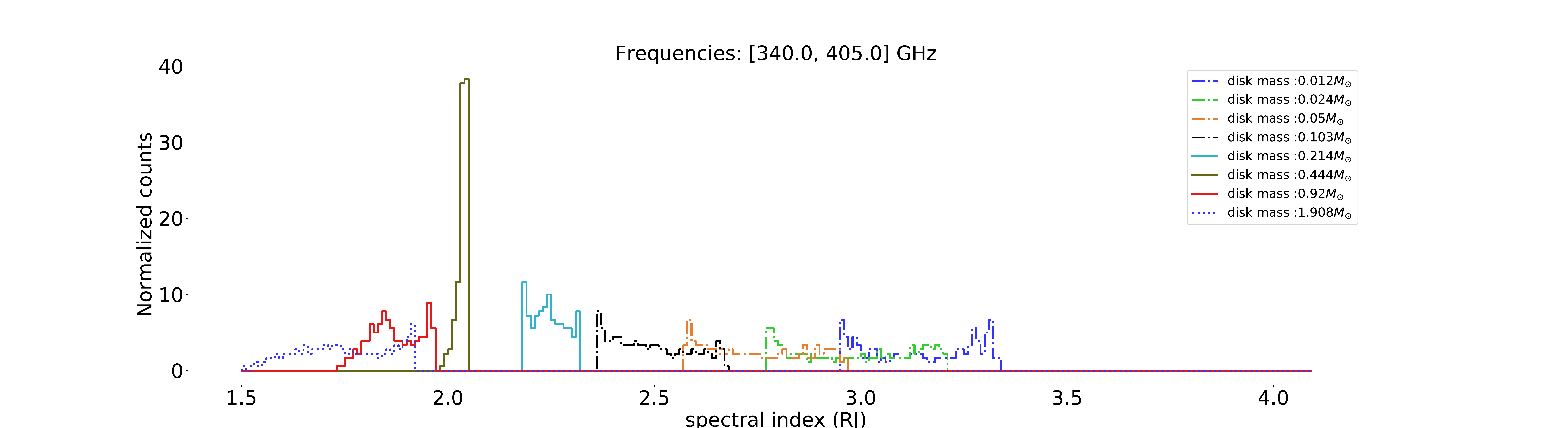}\\
   \end{tabular}
   \caption{Probability distribution function $p_{\alpha_{RJ}}$ of the spectral indices obtained from fitting all ({\it top} panel) and selected frequency bands in the model grid with intrinsic $\beta$=1.7.}
\label{fig:modelalpha_allbands}
\end{figure*}

\begin{figure*}[]
   \begin{tabular}{c}
     \vspace{-0.5cm}\includegraphics[width=\textwidth]{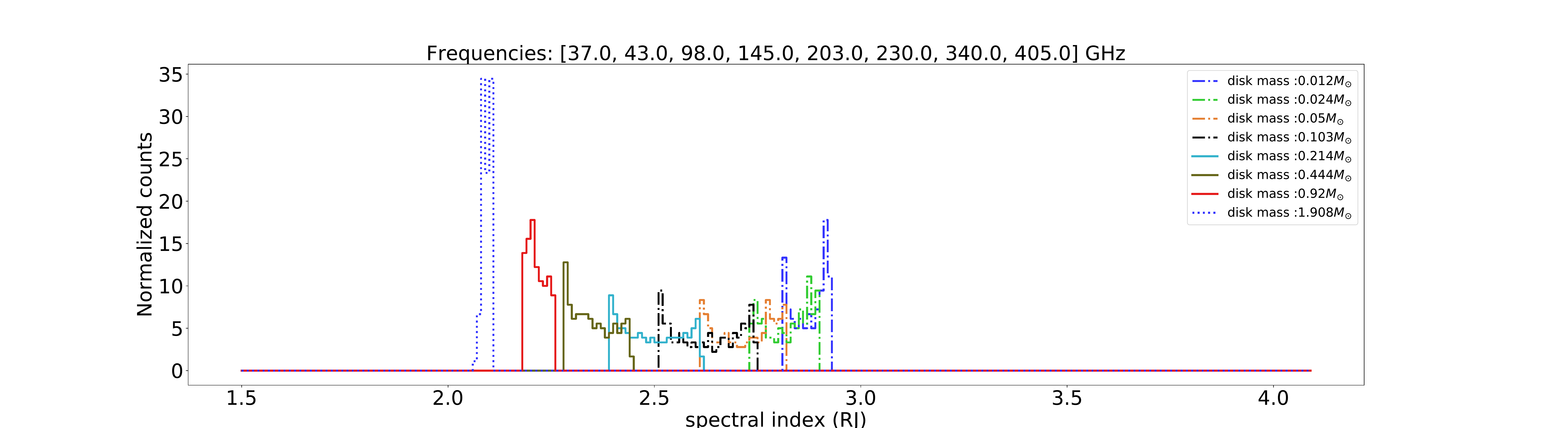}\\
     \vspace{-0.5cm}\includegraphics[width=\textwidth]{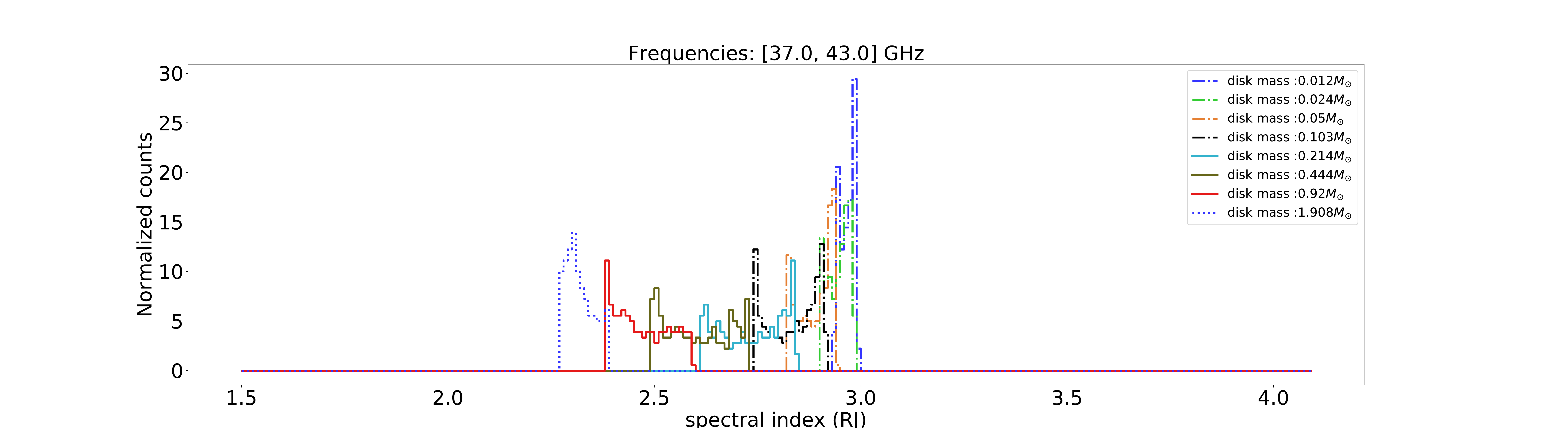}\\
     \vspace{-0.5cm}\includegraphics[width=\textwidth]{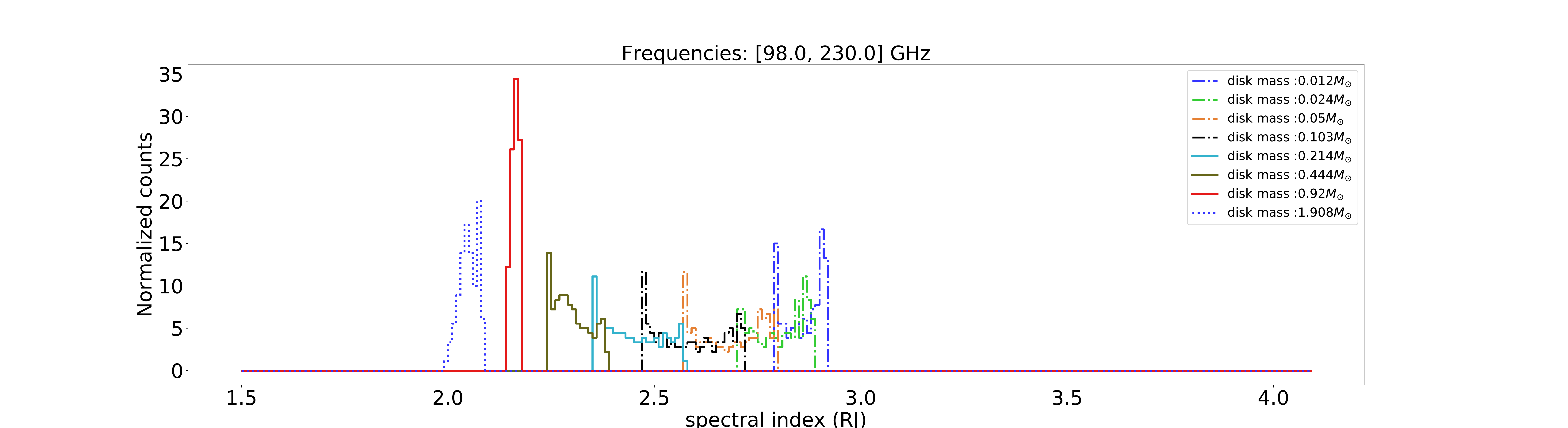}\\
     \vspace{-0.5cm}\includegraphics[width=\textwidth]{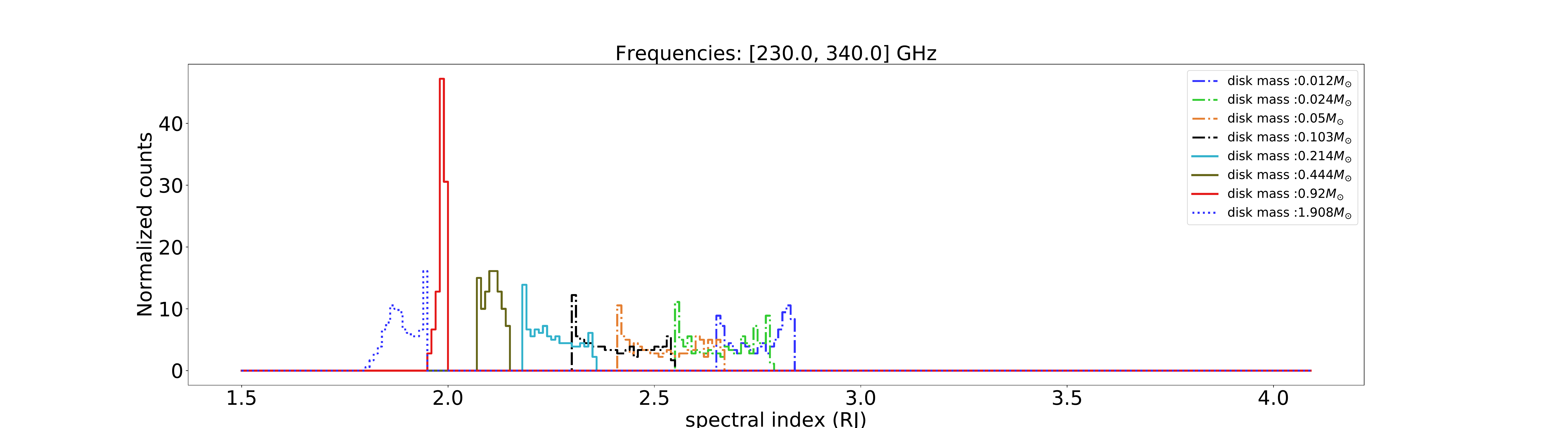}\\
     \vspace{-0.3cm}\includegraphics[width=\textwidth]{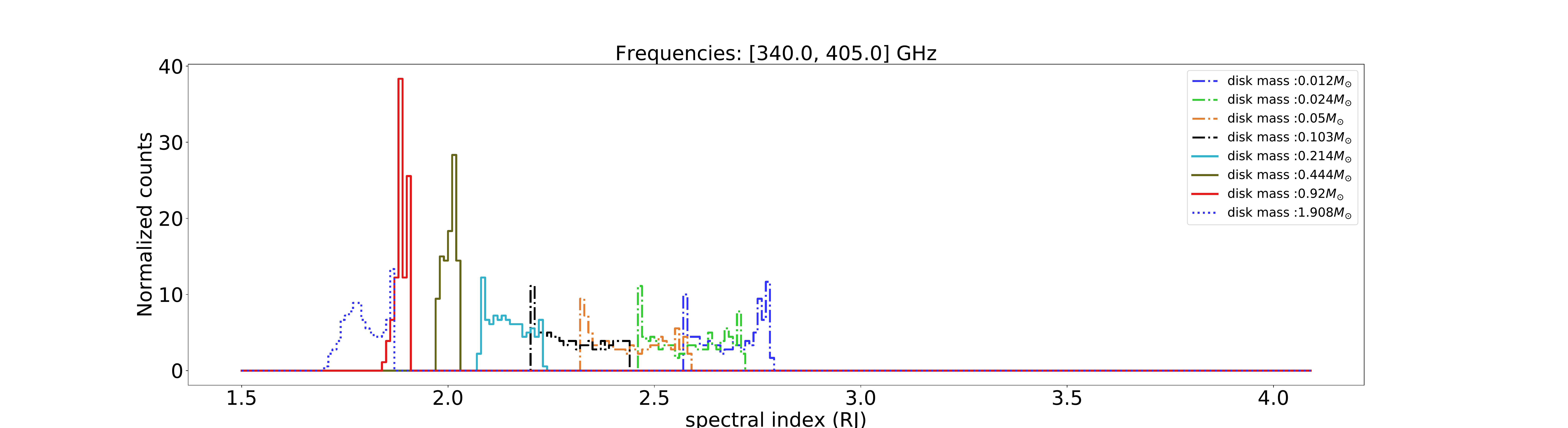}\\
   \end{tabular}
   \caption{Same as in Fig. \ref{fig:modelalpha_allbands}, but for the model grid with intrinsic $\beta$=1.0.}
\label{fig:modelalpha_allbands_beta1p0}
\end{figure*}

For every disk mass and inclination angle, we fit the respective spectral index $\alpha_{RJ}$\footnote{Often referred to as $\beta_{RJ}$+2.0. We prefer to refer to $\alpha_{RJ}$, since one of the main results of this paper is that low $\alpha$'s can be obtained without changing $\beta$.} for different selections of observational frequency bands. Figures \ref{fig:modelalpha_allbands} and \ref{fig:modelalpha_allbands_beta1p0} show the probability distribution functions of the fitted spectral indices $p_{\alpha_{RJ}}$ for every disk mass and for an assumed $\beta = 1.7$ and 1, respectively. The selection of frequencies in each panel of Figs. \ref{fig:modelalpha_allbands} and \ref{fig:modelalpha_allbands_beta1p0} is intended to sample several observational setups varying from rich data sets with many frequency bands, to more modest -- and frequent -- cases where only a couple of bands are available. The histogram of $p_{\alpha_{RJ}}$ for each mass has two peaks at its low- and high-$\alpha_{RJ}$ ends. The high-$\alpha_{RJ}$ peak arises from the fact that the probability of observing a disk close to edge-on is  higher than of observing it close to face-on, whereas the peak at low $\alpha_{RJ}$ is due to the optical depth changing slower with inclination when the disk is closer to face-on (this can also be seen in Fig. \ref{fig:representative_spix}). 

We found that for a fixed disk mass in the range 0.01 to 2 $M_{\odot}$, randomly distributed inclination angles are not enough to broaden  $p_{\alpha_{RJ}}$ to cover the range of values reported by \cite{LiLiu17}. To interpret the widely distributed range of observed $\alpha_{RJ}$ values (Section \ref{sec:obsalpha}), one either needs to assume a broad range of observed disk masses, or a range of dust opacity indices $\beta$, or both. 

Intriguingly, some fits yield values $\alpha_{RJ} < 2$. This happens when the selected frequencies sample the appropriate change in the temperature structure within the model disk. Three conditions need to be fulfilled: 1) having a temperature gradient along the line-of-sight, 2) having large dust optical depths ($\tau >> 1$), and 3) the physical depth sampled by the l.o.s vary significantly with frequency. Thus, the unexpected low values of $\alpha_{RJ}<2.0$ are due to the transition from hotter black body emission at lower frequencies to cooler black body emission at higher frequencies. In our model grids, $\alpha_{RJ} < 2$ occurs only for high masses ($M_{\rm disk} > 0.53~M_\odot$), preferably using combinations of high frequency bands (above Band 6), and at all inclination angles but preferably closer to face-on, in which case the temperature gradient sampled between different bands is larger.  

Spectral indices $< 2$ have indeed been reported several times in the literature \citep{Jorgensen09,Miotello14,LiLiu17,Liu18}. Sometimes they have been attributed to calibration errors, very low dust temperatures \citep[e.g.,][]{Hirano14}, or a significant contribution of free-free emission \citep{Liu17}. In some cases, a very low dust temperature can be ruled out from the observed brightness temperatures \citep{LiLiu17}.

Counterintuitively, a comparison of the bottom two panels of Figure \ref{fig:modelalpha_allbands} and \ref{fig:modelalpha_allbands_beta1p0} shows that $\alpha_{RJ}<2.0$  can be more easily achieved with the 
assumption of $\beta=1.7$ than with $\beta=1.0$. This is due to the faster change of $\tau$ with frequency in the former case.

\subsubsection{Observed probability distribution function of (sub)millimeter spectral indices} \label{sec:obsalpha}

\begin{figure*}[!h]
   \hspace{-1.85cm}
   \begin{tabular}{ p{9cm} p{9cm} }
     \includegraphics[width=11cm]{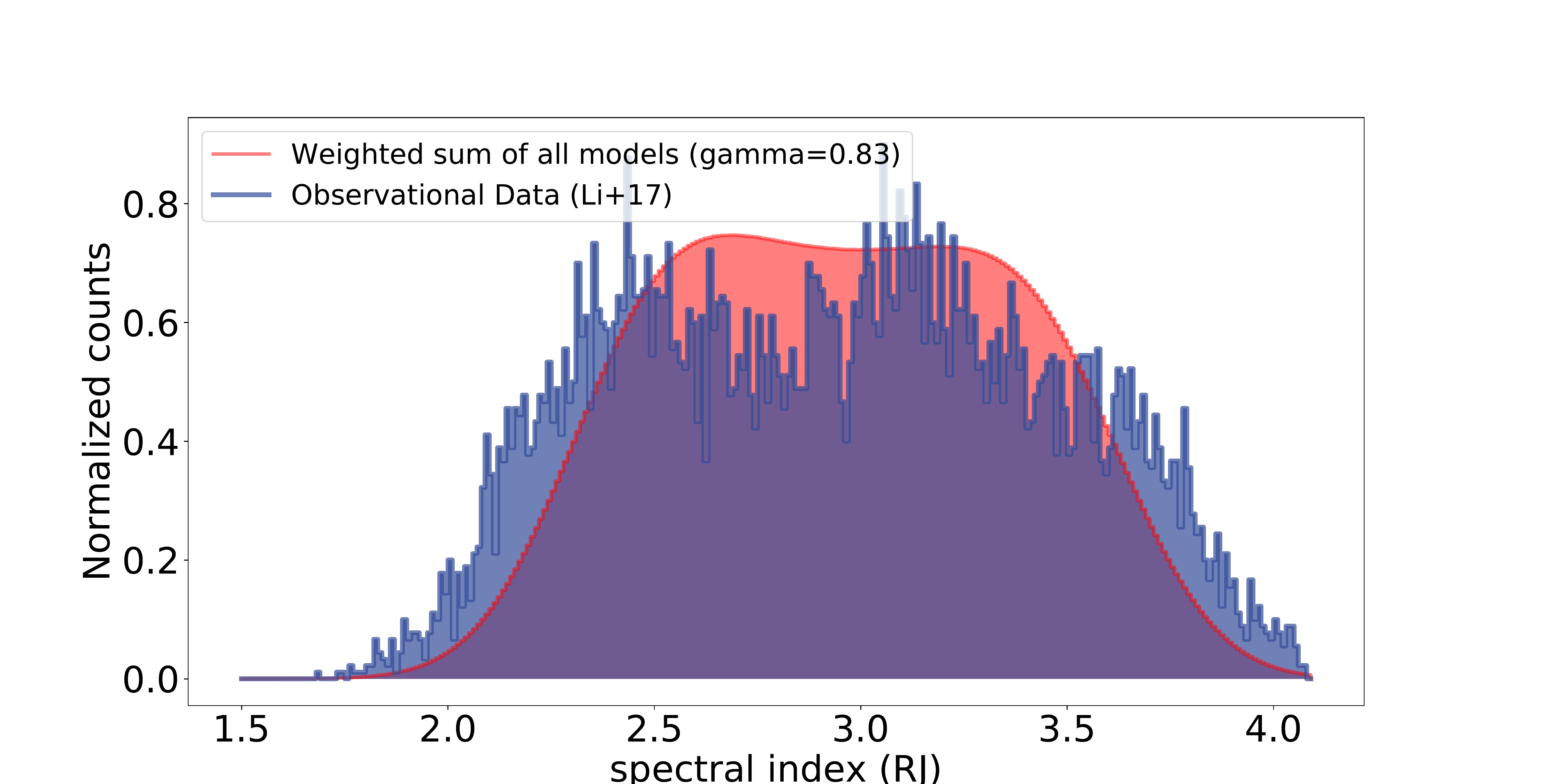} & 
     \includegraphics[width=11cm]{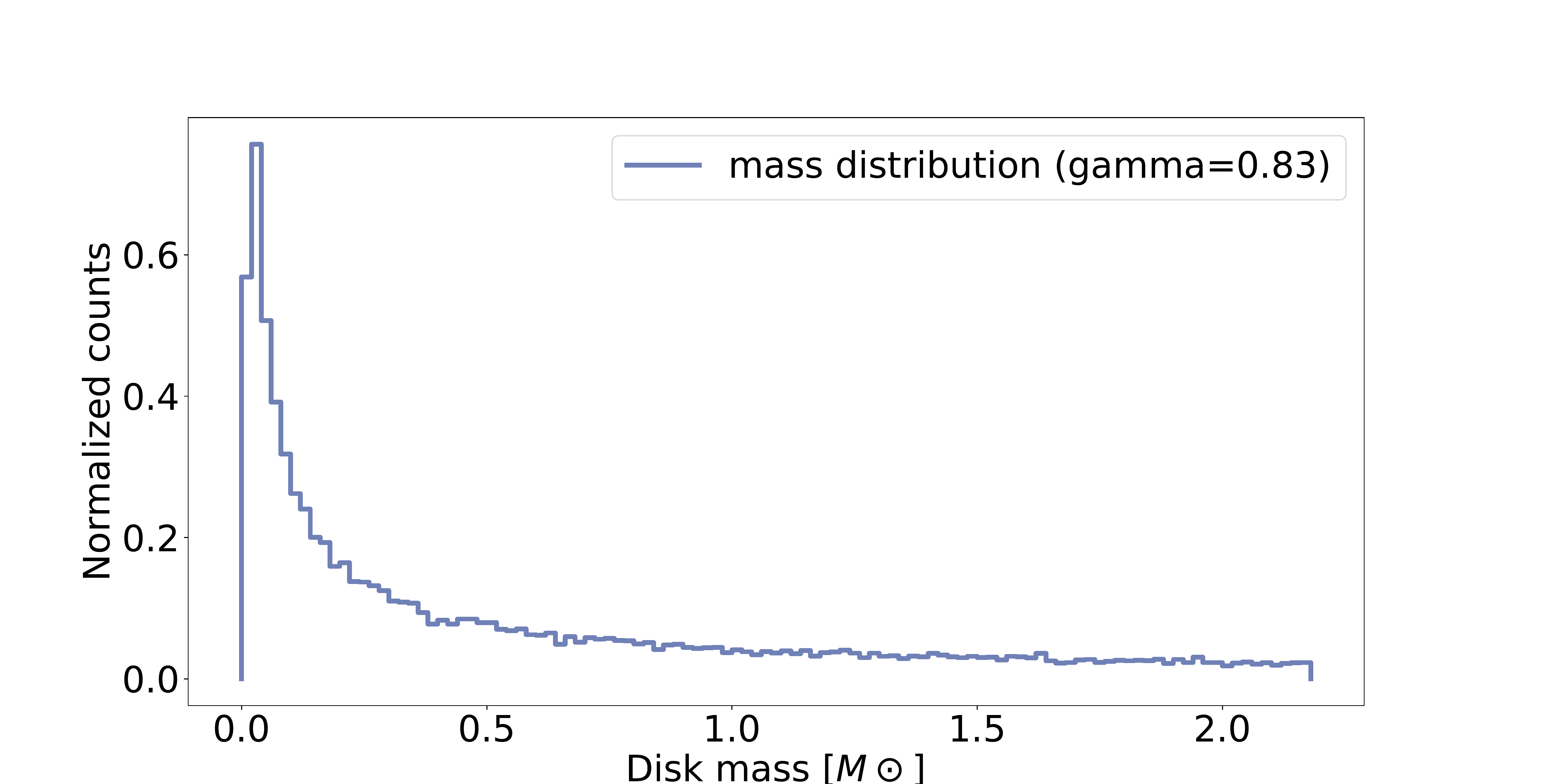} \\
     \includegraphics[width=11cm]{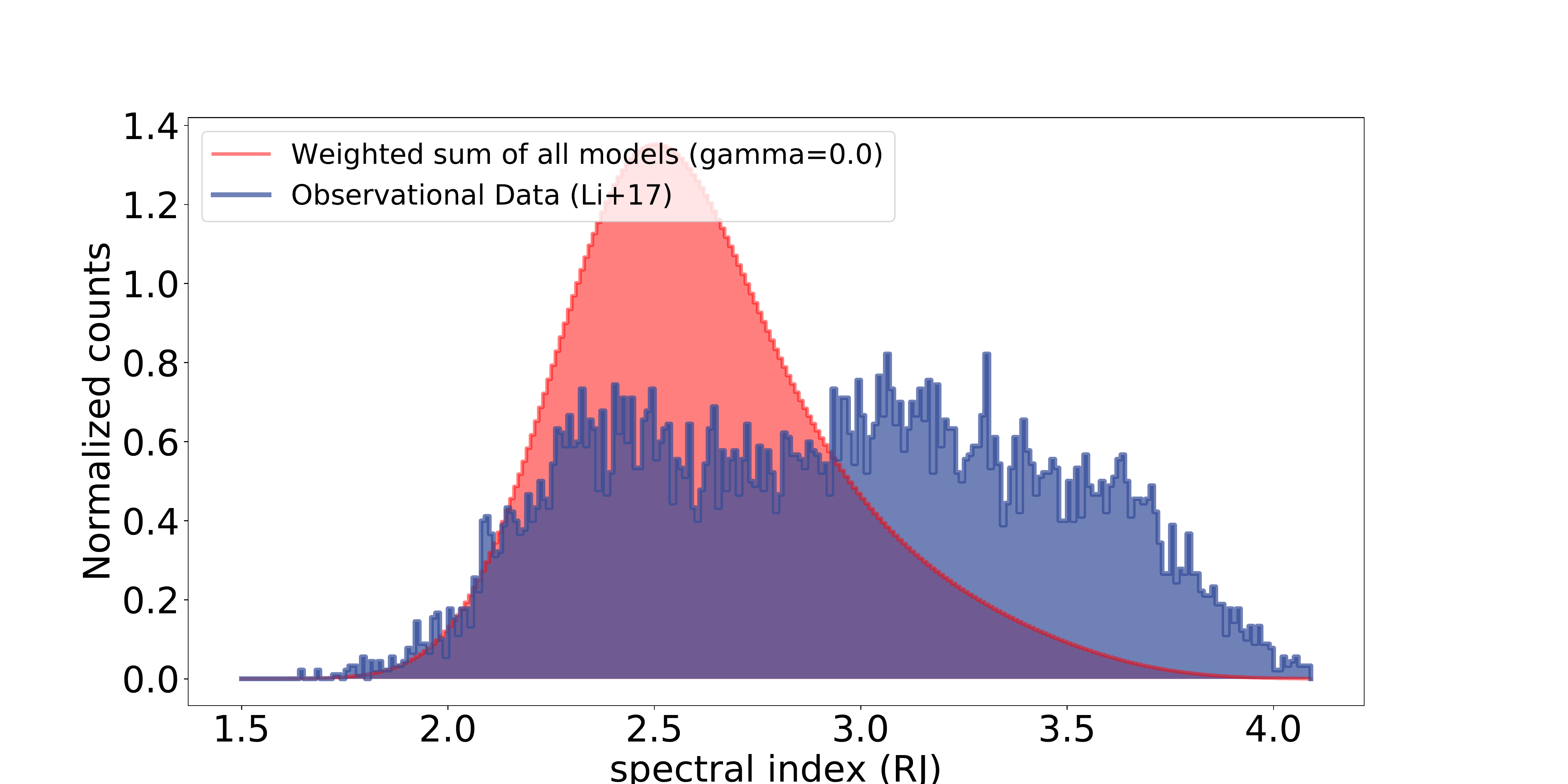} & 
     \includegraphics[width=11cm]{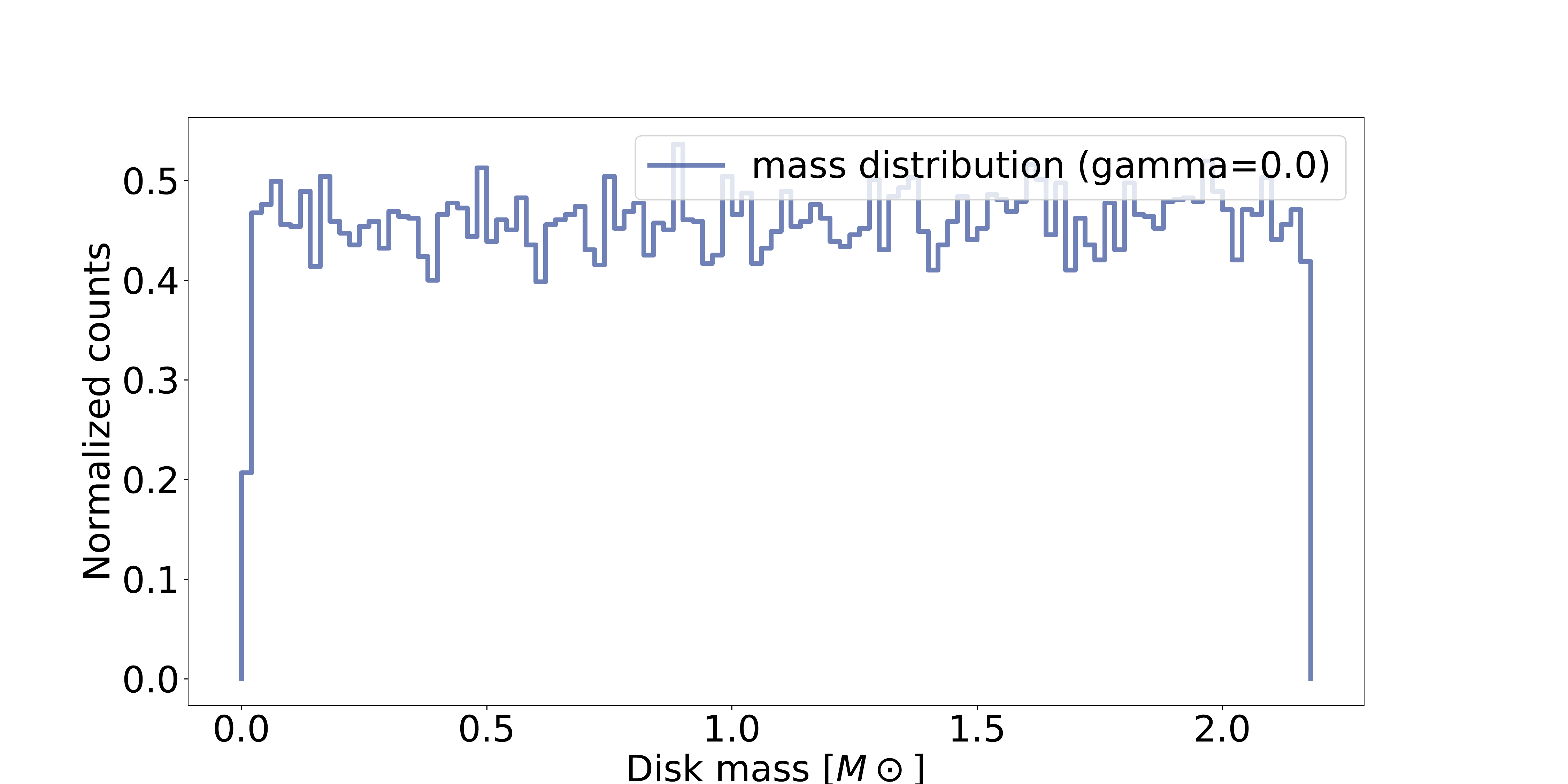} \\
     \includegraphics[width=11cm]{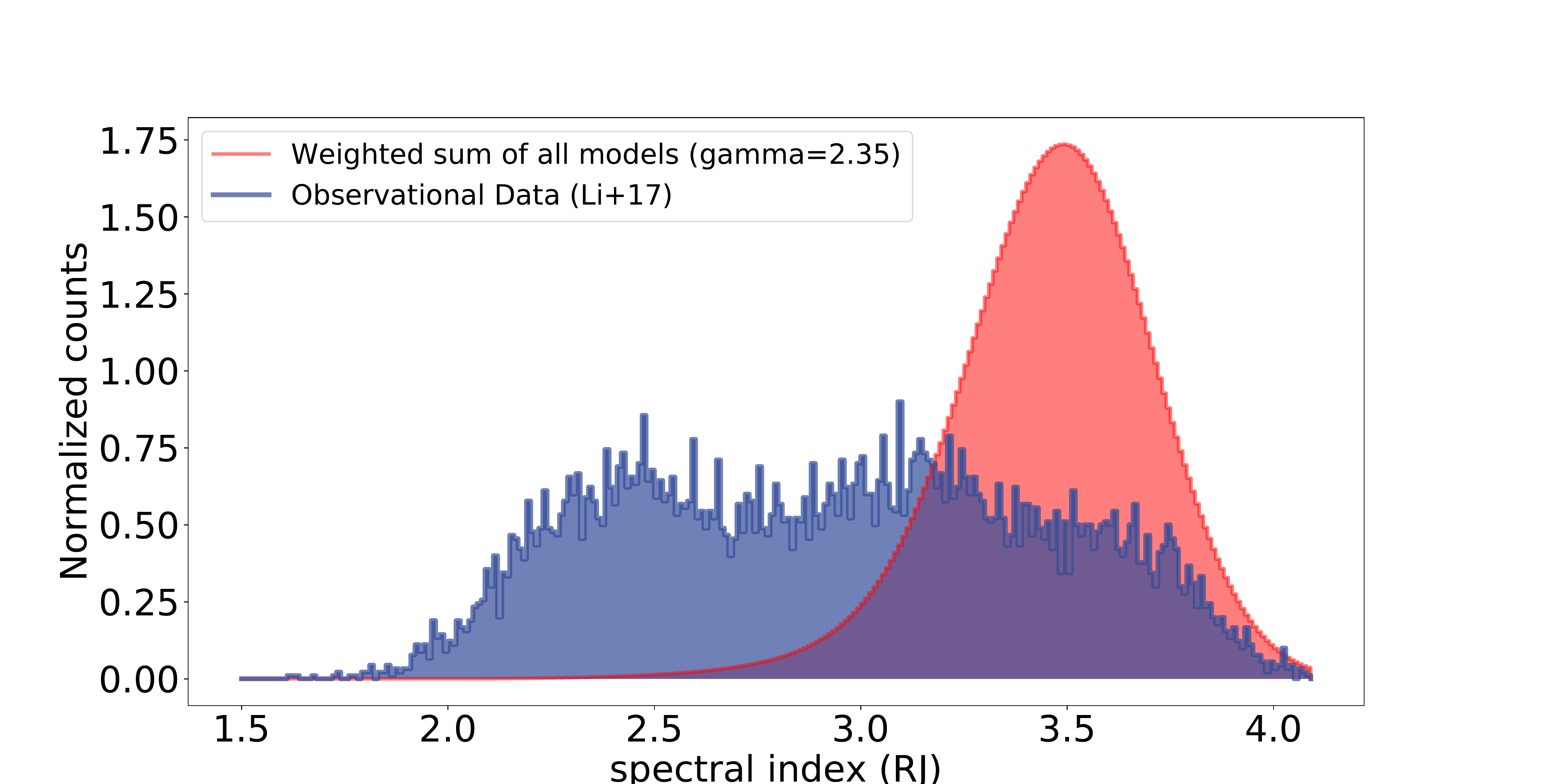} & 
     \includegraphics[width=11cm]{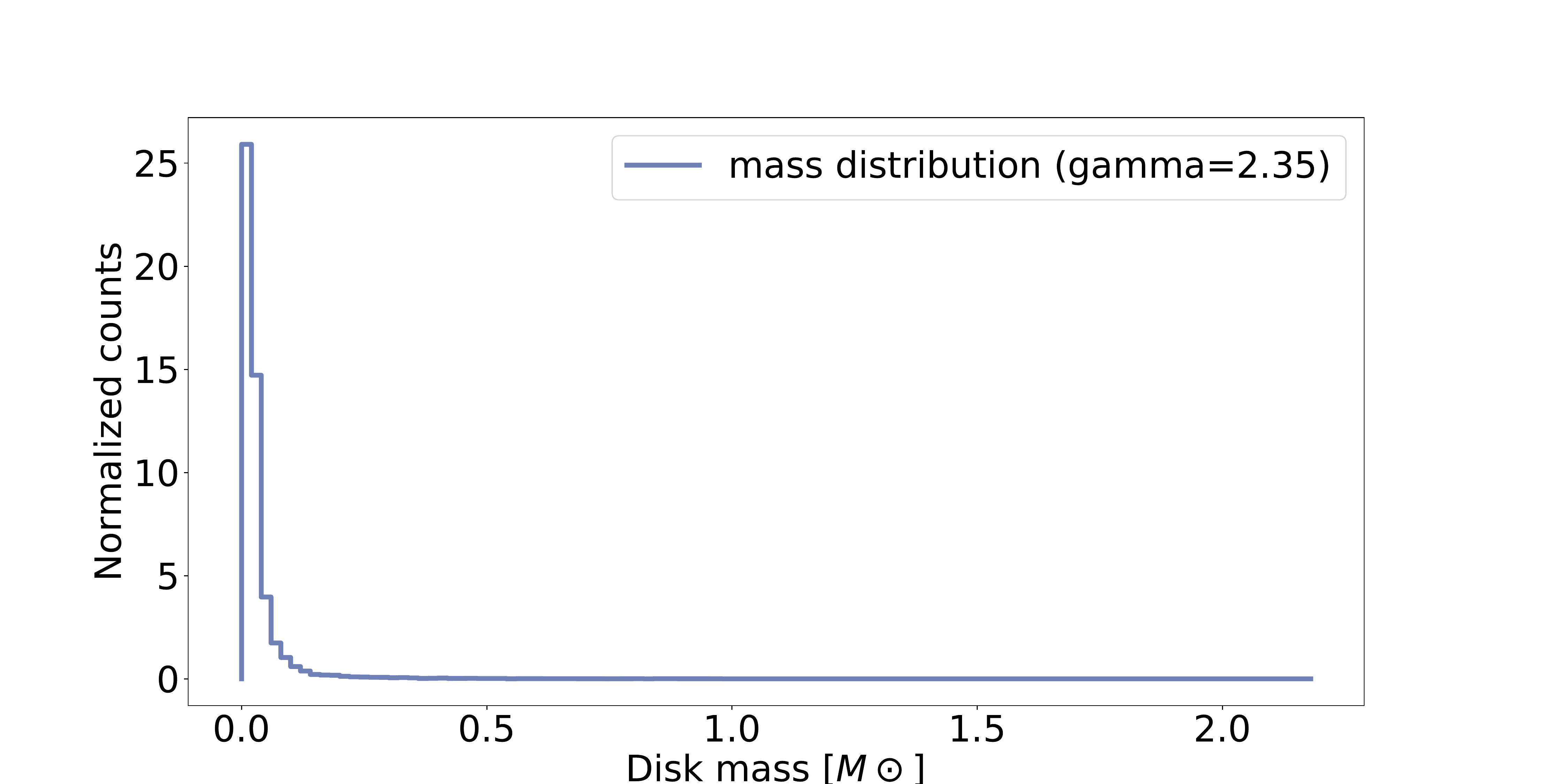} \\
   \end{tabular}
   \vspace{0.2cm}
   \caption{{\it Left} column: probability distribution function of the spectral indices $p_{\alpha_{RJ}}$ obtained from fitting selected bands in the model grids (see Section \ref{sec:comparison}) and from the observations of \cite{LiLiu17}. From {\it top} to {\it bottom} are shown the results of assuming a power-law index of the disk mass distribution functions $\gamma=0.83$, 0.0, and 2.35, respectively (see Section \ref{sec:comparealpha}). We present histograms of our randomly sampled disk masses in the right column. The observational data and the model grid values were resampled assuming a nominal Gaussian spectral index error of 0.2.}
   \label{fig:histograms}
\end{figure*}

In this section and the next we investigate whether or not fixing the intrinsic $\beta$ to 1.7 is enough to reproduce $p_{\alpha_{RJ}}$ as derived from the observations reported by \citet{LiLiu17}.

We take from \cite{LiLiu17} the best Rayleigh-Jeans fits for the (sub)millimeter dust opacity indices ($\beta_{RJ}$) for nine Class 0/I YSOs, namely: L1448\,CN, NGC\,1333 IRAS2A, NGC\,1333 IRAS4A1, NGC\,1333 IRAS4A2, Barnard\,1b-N, Barnard\,1b-S, L1527, Serpens\,FIRS1, and L1157. From the fitted $\beta_{RJ}$, we define $\alpha_{RJ} = \beta_{RJ}+2$ and work with this quantity. 

To have a more fair comparison with our grids of models (Section \ref{sec:grid_of_models}), we redo the fits for Barnard\,1b-N and Barnard\,1b-S without including the 280 GHz measurements.
We obtain $\alpha_{RJ}=$3.19 and $\alpha_{RJ}=$3.65, respectively, both in good agreement with the original fits. From a comparison of the best fit and median values given in Table 4 of \cite{LiLiu17}, we estimate a nominal Gaussian measurement error  $\sigma(\alpha_{RJ})=0.2$. 
We then re-sample $\alpha_{RJ}$ for each of the nine sources 1000 times by adding the estimated noise, and produce an approximated $\alpha_{RJ}$ probability distribution function (PDF) $p_{\alpha_{RJ}}$ from the 9000 samples of $\alpha_{RJ}$.
We note that we cannot infer $p_{\alpha_{RJ}}$ based on a Bayesian approach, given that our observational sample size is small and that it is not clear how to make an appropriate assumption about the a priori distributions of the parameters. 
Our present approach for obtaining $p_{\alpha_{RJ}}$ is a compromise in this sense, and thus the $p_{\alpha_{RJ}}$ that we obtain may be broader than the actual PDF by $\sim\sigma(\alpha_{RJ})$.
Nevertheless, the present error in $p_{\alpha_{RJ}}$ is likely dominated by stochastic effects due to the small observational sample, and not by the way in which way we determine  $\sigma(\alpha_{RJ})$ and $p_{\alpha_{RJ}}$.\\

\subsubsection{Comparing weighted averages of models to observations}\label{sec:comparealpha} 

As mentioned in Section \ref{sec:gridalpha}, we consider 180 inclination angles $i$ which uniformly sample $\cos (i)$ in the range [0, 1].
In addition, we produced $3\times10^4$ random samples of disk masses assuming that the probability to find a disk with mass between $M$ and $M+dM$, $P(M)$, is proportional to $M^{-\gamma}$. 
We explored several cases between $\gamma=2.35$ and $\gamma=0.0$, and consider the case with  $\gamma=0.83$  (with a nominal error of $\sim\pm$0.05) as our fidicual value to compare to observations. 
The choice of $\gamma=2.35$ was motivated by the stellar initial mass functions (IMF) proposed by \citet{Salpeter55}, while the choice of $\gamma=0.0$ (flat mass distribution) provides another extreme for comparison.
We adopted lower and upper disk mass cutoffs at 0.012 $M_{\odot}$ and 2.290 $M_{\odot}$, respectively.
We note that for our fiducial case and for $\gamma=2.35$, the exact value for the upper cutoff is not important since the cumulative  probabilities for disks at $>2$ $M_{\odot}$ are very low.
Our present lower mass cutoff is realistic considering the typical detection limits of the measurements in \cite{LiLiu17}.

To mimic the observations of the nine Class 0/I sources reported by \cite{LiLiu17}, we fit $\alpha_{RJ}$ for each of the 
180$\times$30000 interpolated models using the following six sets of frequencies in GHz: [37, 43, 230, 340], [37, 43, 98, 203], [37, 230, 340], [37, 98, 230, 340], [43, 98, 230, 340], and [43, 230, 340]. We then averaged the fits from these six sets of frequencies assuming a relative weighting of [2.0, 1.0, 2.0, 1.0, 1.0, 2.0], as was the case for the  observed sources in \cite{LiLiu17}. 
Similarly to what we did to produce $p_{\alpha_{RJ}}$ from the observations, we re-sampled 100 times the derived $\alpha_{RJ}$ for each of the individual 180$\times$30000 models, by adding the observational $\sigma(\alpha_{RJ})=0.2$ Gaussian random noise, and then evaluated an approximated $p_{\alpha_{RJ}}$ from these 180$\times$30000$\times$1000 samples.

The left column of Figure \ref{fig:histograms} shows the comparison of $p_{\alpha_{RJ}}$ evaluated from the interpolated grids of $\beta=$1.7 models with that evaluated from the observations (Section \ref{sec:obsalpha}).
The right column presents histograms of our randomly sampled disk masses.
We found that $p_{\alpha_{RJ}}$ for the flat disk-mass distribution $\gamma=0.0$ has an excess of low $\alpha_{RJ}$ values compared to observations. In contrast, $p_{\alpha_{RJ}}$ for $\gamma=2.35$ presents an excess of high $\alpha_{RJ}$ values. 
A $\gamma$ value in between these two extreme cases, such as our fiducial $\gamma=0.83$, can naturally explain the observed $p_{\alpha_{RJ}}$.
From figure \ref{fig:modelalpha_allbands} we can see that when assuming an intrinsic $\beta\sim$1.7 for a broad range of disk masses, the variation of $\alpha_{RJ}$ is approximately proportional to the variation of $\log(M_{\mbox{\scriptsize disk}})$.
To yield a rather uniform sampling of $\alpha_{RJ}$ as in the observational $p_{\alpha_{RJ}}$ thus requires the value of $\gamma$ to be close to 1.0.

\begin{figure}[]
   \hspace{-1.85cm}
   \begin{tabular}{ p{9cm} }
     \includegraphics[width=11cm]{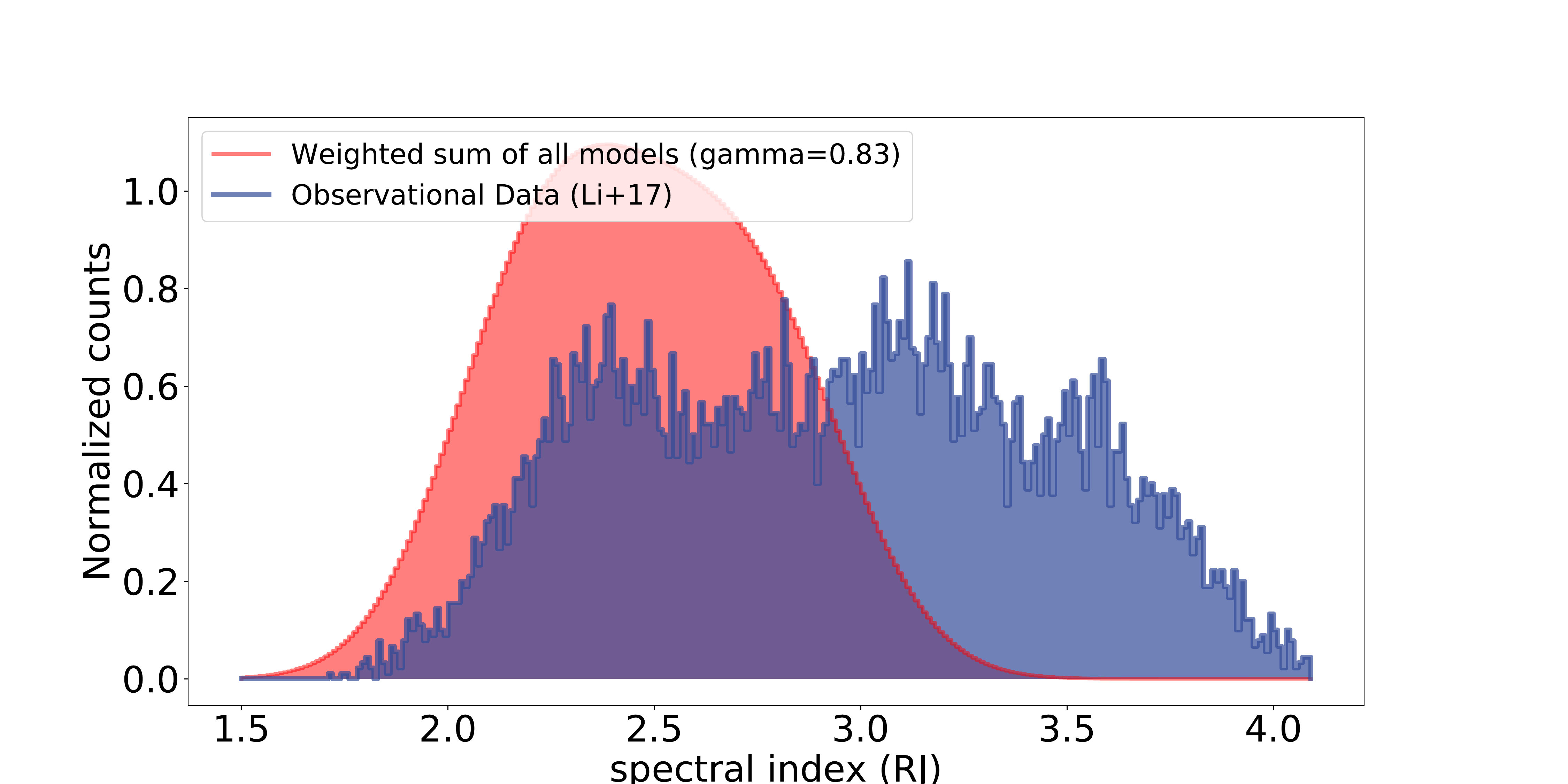}  \\
     \includegraphics[width=11cm]{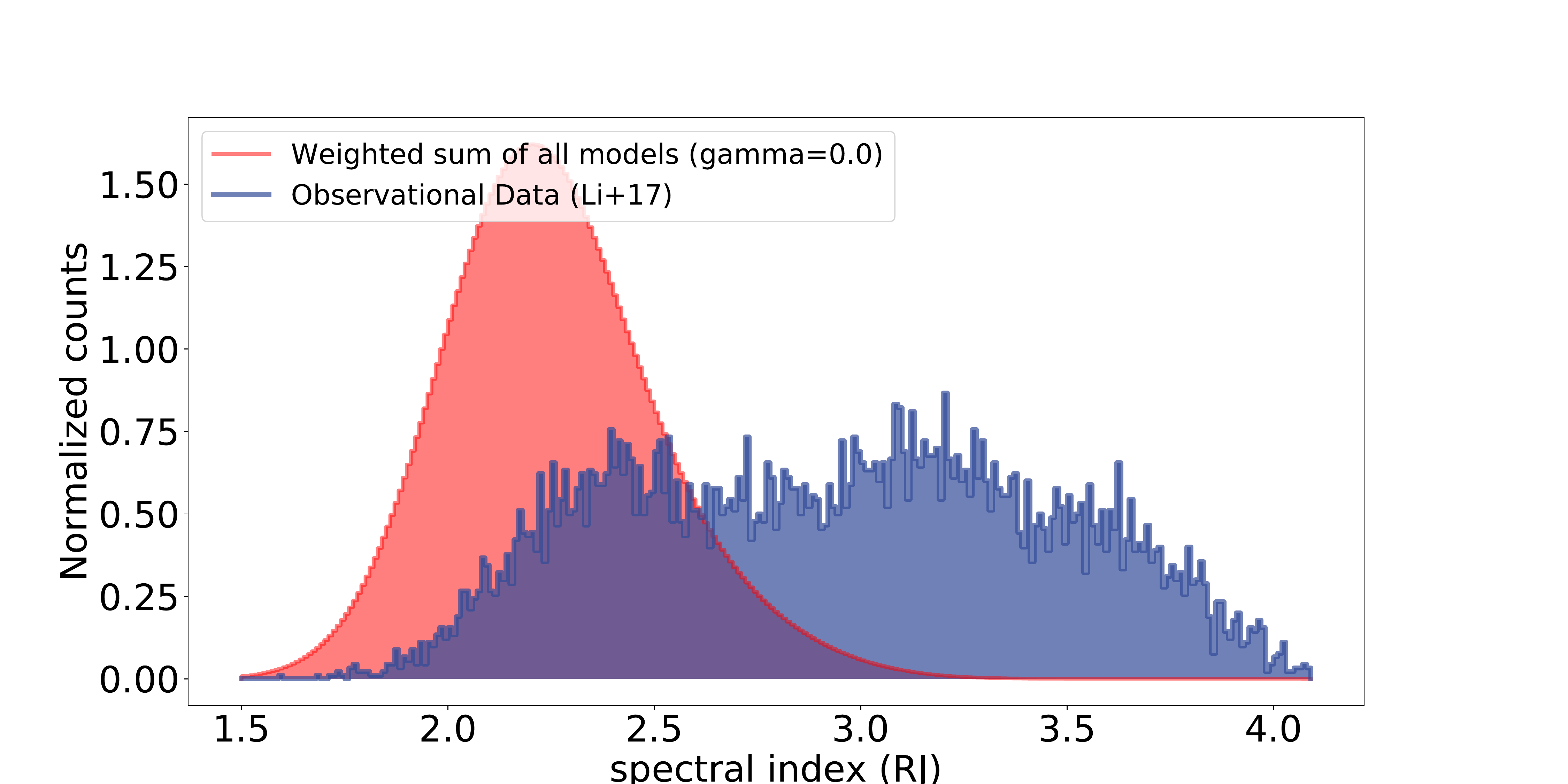}  \\
     \includegraphics[width=11cm]{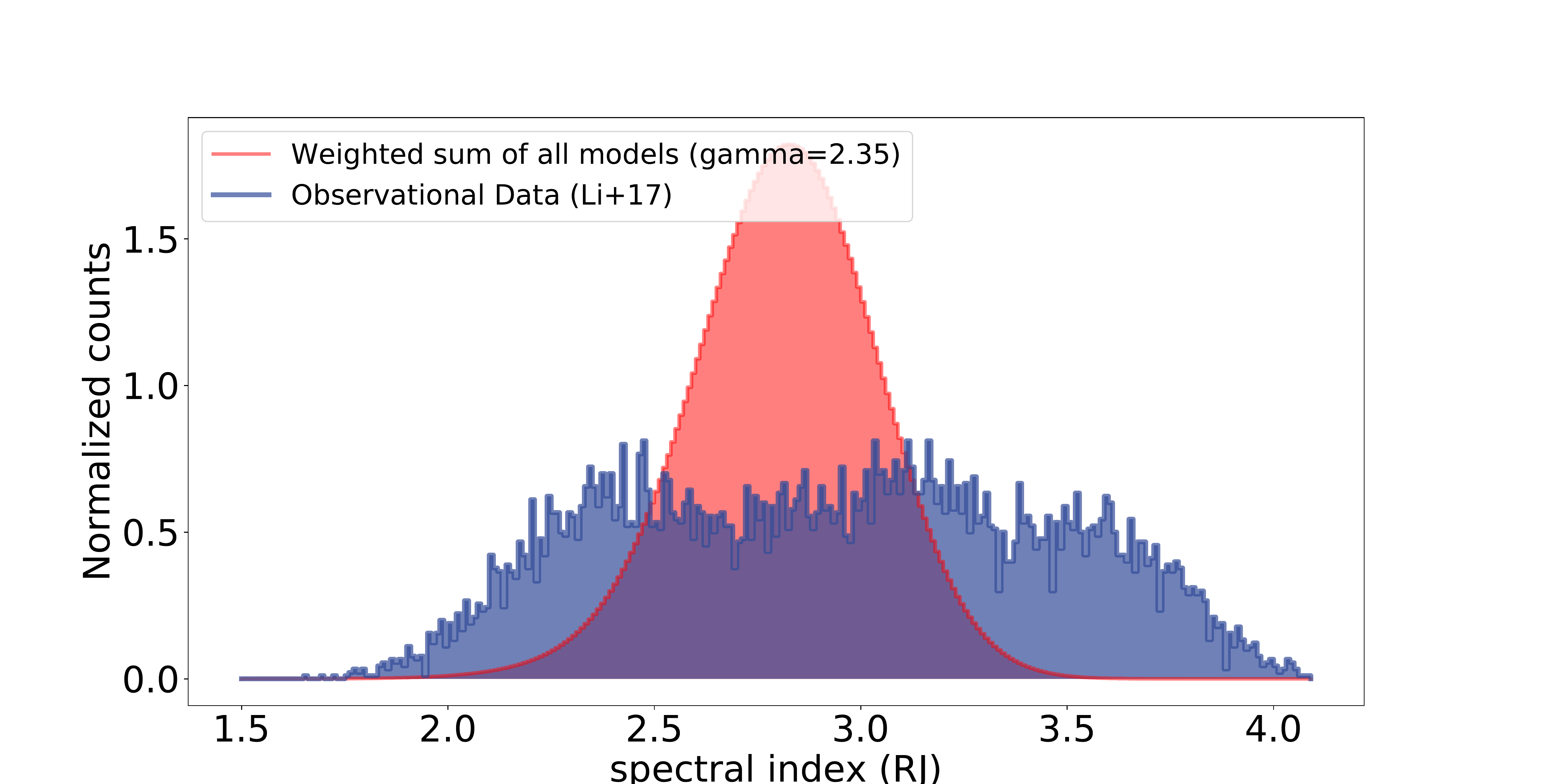}  \\
   \end{tabular}
   \vspace{0.2cm}
   \caption{Same as in the left column of figure \ref{fig:histograms}, but for the model grid with an intrinsic $\beta$=1.0.}
   \label{fig:histograms_beta1p0}
\end{figure}

On the other hand, figure \ref{fig:histograms_beta1p0} shows that assuming a constant $\beta=$1.0 makes $p_{\alpha_{RJ}}$ to be too biased toward low values compared to observations.
With the assumption of $\beta=$1.0, to explain the population of the observed sources with $\alpha_{RJ}\sim$3.5, it is needed to artificially enlarge the assumption of measurement errors which will however also produce too many sources at unrealistically low $\alpha_{RJ}$.

\section{Discussion} \label{sec:discussion}

\subsection{On the origin of low spectral indices} \label{sec:disc_low_alpha}

We were able to reproduce the observed distribution of dust spectral indices in the protostellar sample of \cite{LiLiu17} {\it without} invoking a decrease in the (sub)mm dust opacity power-law exponent $\beta$ from the typical ISM value of 1.7. This does not mean that {\it some} grain growth cannot occur in protostellar disks, but it certainly suggests that there is no need to invoke dust evolution to the levels inferred in class II samples \citep{Testi14} within the $\times 10$ shorter protostellar lifetimes as compared to class II lifetimes \citep{Dunham14}. 

Recent (sub)mm polarization studies have shown that the properties of self-scattered emission can be modelled to infer the degree of grain growth \citep{Kataoka15}. Quantitative determinations of maximum grain sizes using this technique have found $a_\mathrm{max} \sim 60-150~\mu$m for class I and II YSOs \citep[e.g.,][]{Yang16,Kataoka16,Hull18}, which seems in `tension' with grain growth as inferred from spectral indices\footnote{For a maximum grain size $\sim 100~\mu$m, the expected $\beta_\mathrm{0.8mm-9mm}$ is not smaller than 1.5, regardless of composition, see, e.g., \cite{Testi14}.}. However, it is not clear to which extent self-scattering can dominate over magnetic-field alignment in the polarized emission \citep[e.g.,][]{Lee18,Girart18}. It is possible that in class 0 and I protostars magnetic fields tend to dominate \citep{Alves18,Liu18}.   
We conclude that self-obscuration due to relatively high masses and temperature structure could be an important part of the puzzle, helping to explain low spectral indices without needing to invoke significant dust evolution at the earliest YSO stages.  

Indeed, we found that when a temperature gradient is present and there are large changes of effective optical depth with frequency, the R-J spectral index measured from the models can be  $\alpha_{RJ}<2.0$, as it has been reported for some objects. 
These observational measurements, if not due to calibration errors, 
cannot be reproduced by decreasing the intrinsic $\beta$ values alone, but need the above mentioned conditions for objects with a range of masses and inclination angles. 

The way in which $p_{\alpha_{RJ}}$ is sensitive to the assumption of the power law $\gamma$ for an underlying `protostellar disk mass function' is intriguing. A value $\gamma=0.83$ matches well current observations, and it resembles the power-law index of the \citet{Miller1979} or \citet{Kroupa01} IMF in the specific stellar mass range of $\sim$0.01-1 $M_{\odot}$. However, any relation 
between the class 0/I disk masses and the final stellar mass is far from clear. Also, the observations of \cite{LiLiu17} may be subject to target selection biases. Future surveys of larger samples of protostars may provide more robust conclusions on this topic. 

\subsection{Massive protostellar disks and implications \\ for planet formation} \label{sec:planet_formation}

Our results show that a range of protostellar disk masses $M_{\rm disk} \sim 0.01-2~M_\odot$ is needed to explain the wide observed range of (sub)mm spectral indices. Masses $> 0.1~M_\odot$ are larger than many determinations in the literature, which often assume optically-thin dust emission (see Section \ref{sec:intro}) and/or (sub)mm dust opacities $\kappa_\nu$ well above our selection of $\kappa_{\rm 230GHz} = 1$ cm$^2$ g$^{-1}$, which we consider to be already close to the high end of valid values. 
Often, higher (sub)mm opacities are used (and lower masses are obtained) because the normalization of the seminal paper by \cite{Beckwith90} at 1 THz is used with $\beta < 1$. However, we think that this selection is questionable because the normalization $\kappa_0$ and exponent $\beta$ are not really independent from each other \citep[see, e.g.,][]{Dalessio01,Draine06}. To get $\beta$ as small as $\sim 0$, it requires the maximum grain size to be $>> 1$ cm, in which case dust cannot emit efficiently at (sub)mm wavelengths, i.e., the value of $\kappa_\nu$ at these wavelengths ought to be small, and the derived masses are therefore large compared to a high-$\kappa$ assumption. 
In summary, we believe that our result on the need for {\it some} protostellar disks significantly more massive than $0.1~M_\odot$ is a robust one. 

Without considering the complicated physics of dust fragmentation and inward migration \citep[e.g.,][]{Birnstiel10}, one may naively expect that the more massive protostellar disks would have more efficient grain growth due to higher collision rates.
If this is approximately true, then a more realistic model assumption may be to allow the lower mass disks to have $\beta\sim$1.7 and the higher mass ones to have smaller $\beta$.
Such an assumption would yield a $p_{\alpha_{RJ}}$ broader than that presented in the top row of figure \ref{fig:histograms} ($\beta=$1.7, $\gamma=$0.83).
Therefore, our present fiducial $\gamma$ may be considered as a lower limit.

A more realistic dependence of $\beta$ with disk mass could be provided from hydrodynamic simulations which self-consistently evaluate dust grain growth and migration \citep[e.g.,][]{Vorobyov18}.
Also, better observational constraints will come from current and future ALMA and VLA surveys of protostellar disks. 

Finally, if protostellar disk masses are systematically larger than often thought, it is also possible that the initial seeds of planet formation are already present since the class 0/I stage \citep[e.g.,][]{Zhu12}. Such scenario would help to solve the apparent mass-budget problem currently being debated for the more evolved class II protoplanetary disks, which often appear to have less mass than the necessary to form giant planets \citep[e.g.,][]{Ansdell16,Miotello17}. Massive disks can be gravitationally unstable \citep[e.g.,][]{Evans17,Vorobyov18}, but the development of instabilities such as spiral arms does not preclude the formation of a rotationally-supported disk-like structures inside the centrifugal barrier, aided by turbulence, non-ideal MHD effects, and dust chemistry \citep[e.g.,][]{Seifried15,LizanoGalli15,Zhao16,Zhao18}.

\section{Conclusions} \label{sec:conclusions}
\noindent
- We have shown that self-obscuration in protostellar disks with an appropriate temperature gradient has important effects in their appearance. The most striking feature is the existence of dark lanes, which, as expected, become more prominent with increasing mass, inclination angle, and observing frequency. \\
- Self-obscuration also has important effects in the measured spectral indices, making them significantly smaller than the typical value for optically-thin ISM dust $\alpha_{\rm ISM} \approx 3.7$. From our model grids, we 
conclude that spectral indices between all ALMA bands are significantly affected by self-obscuration for all models with $M_{\rm disk} \gtrsim 0.05$, regardless of inclination. \\
- Determinations of protostellar disk masses could often be underestimated by $> \times 10$. Sensitive, high-angular resolution continuum observations at frequencies $\nu < 50$ GHz may be crucial to truly unveil the density structure and mass budget of protostellar disks in optically-thin dust emission. ALMA Band 1 \citep{DiFrancesco13} and a future Next Generation VLA \citep{Isella15} would be the main facilities to achieve this. \\
- From a comparison of the inferred probability distribution functions of spectral indices $p_{\alpha_{RJ}}$ as derived from our model grids and the available observations, 
we find that a distribution of disk masses in the range $M_{\rm disk} = 0.01-2~M_\odot$ is needed to reproduce the data, and that assuming a fixed $\beta =1.7$ gives better results than $\beta=1$. However, some level of dust evolution cannot be discarded during the short protostellar timescales. \\  
- A wide disk mass distribution is also needed to produce cases with $\alpha <2$, as has been reported for some sources in the literature. Such extremely low spectral indices arise when the selected observing frequencies sample the appropriate change in the temperature structure of the optically thick disk.


\acknowledgments
The authors thank the anonymous referee for useful reports.
RGM acknowledges support from UNAM-PAPIIT program IA102817, IRyA-UNAM, and ESO. 
AI thanks the support from Facultad de Ciencias Exactas y Naturales (UdeA), Relaciones Internacionales (UdeA), and IRyA-UNAM. AM and HBL acknowledge support from the ESO Fellowship programme. 
SL acknowledges financial support from CONACyT 238631 and DGAPA-UNAM IN101418. 

\software{SF3dmodels \citep{Izquierdo18}, LIME \citep{Brinch10}, NumPy \citep{Numpy}, APLpy \citep{AplPy}}

\bibliographystyle{yahapj}
\bibliography{references}

\begin{thebibliography}{}
\providecommand\natexlab[1]{#1}
\providecommand\JournalTitle[1]{#1}

\bibitem[{{Alves} {et~al.}(2018){Alves}, {Girart}, {Padovani}, {Galli},
  {Franco}, {Caselli}, {Vlemmings}, {Zhang}, \& {Wiesemeyer}}]{Alves18}
{Alves}, F.~O., {Girart}, J.~M., {Padovani}, M., {et~al.} 2018,
  \href{http://dx.doi.org/10.1051/0004-6361/201832935}{\JournalTitle{\aap},
  616, A56}

\bibitem[{{Ansdell} {et~al.}(2016){Ansdell}, {Williams}, {van der Marel},
  {Carpenter}, {Guidi}, {Hogerheijde}, {Mathews}, {Manara}, {Miotello},
  {Natta}, {Oliveira}, {Tazzari}, {Testi}, {van Dishoeck}, \& {van
  Terwisga}}]{Ansdell16}
{Ansdell}, M., {Williams}, J.~P., {van der Marel}, N., {et~al.} 2016,
  \href{http://dx.doi.org/10.3847/0004-637X/828/1/46}{\JournalTitle{\apj}, 828,
  46}

\bibitem[{{Bate}(2018)}]{Bate18}
{Bate}, M.~R. 2018,
  \href{http://dx.doi.org/10.1093/mnras/sty169}{\JournalTitle{\mnras}, 475,
  5618}

\bibitem[{{Beckwith} {et~al.}(1990){Beckwith}, {Sargent}, {Chini}, \&
  {Guesten}}]{Beckwith90}
{Beckwith}, S.~V.~W., {Sargent}, A.~I., {Chini}, R.~S., \& {Guesten}, R. 1990,
  \href{http://dx.doi.org/10.1086/115385}{\JournalTitle{\aj}, 99, 924}

\bibitem[{{Bergin} {et~al.}(2013){Bergin}, {Cleeves}, {Gorti}, {Zhang},
  {Blake}, {Green}, {Andrews}, {Evans}, {Henning}, {{\"O}berg}, {Pontoppidan},
  {Qi}, {Salyk}, \& {van Dishoeck}}]{Bergin13}
{Bergin}, E.~A., {Cleeves}, L.~I., {Gorti}, U., {et~al.} 2013,
  \href{http://dx.doi.org/10.1038/nature11805}{\JournalTitle{\nat}, 493, 644}

\bibitem[{{Birnstiel} {et~al.}(2010){Birnstiel}, {Dullemond}, \&
  {Brauer}}]{Birnstiel10}
{Birnstiel}, T., {Dullemond}, C.~P., \& {Brauer}, F. 2010,
  \href{http://dx.doi.org/10.1051/0004-6361/200913731}{\JournalTitle{\aap},
  513, A79}

\bibitem[{{Brinch} \& {Hogerheijde}(2010)}]{Brinch10}
{Brinch}, C., \& {Hogerheijde}, M.~R. 2010,
  \href{http://dx.doi.org/10.1051/0004-6361/201015333}{\JournalTitle{\aap},
  523, A25}

\bibitem[{{Carrasco-Gonz{\'a}lez} {et~al.}(2016){Carrasco-Gonz{\'a}lez},
  {Henning}, {Chandler}, {Linz}, {P{\'e}rez}, {Rodr{\'{\i}}guez},
  {Galv{\'a}n-Madrid}, {Anglada}, {Birnstiel}, {van Boekel}, {Flock}, {Klahr},
  {Macias}, {Menten}, {Osorio}, {Testi}, {Torrelles}, \& {Zhu}}]{CG16}
{Carrasco-Gonz{\'a}lez}, C., {Henning}, T., {Chandler}, C.~J., {et~al.} 2016,
  \href{http://dx.doi.org/10.3847/2041-8205/821/1/L16}{\JournalTitle{\apjl},
  821, L16}

\bibitem[{{Chiang} {et~al.}(2012){Chiang}, {Looney}, \& {Tobin}}]{Chiang2012}
{Chiang}, H.-F., {Looney}, L.~W., \& {Tobin}, J.~J. 2012,
  \href{http://dx.doi.org/10.1088/0004-637X/756/2/168}{\JournalTitle{\apj},
  756, 168}

\bibitem[{{Choi} {et~al.}(2010){Choi}, {Tatematsu}, \& {Kang}}]{Choi2010}
{Choi}, M., {Tatematsu}, K., \& {Kang}, M. 2010,
  \href{http://dx.doi.org/10.1088/2041-8205/723/1/L34}{\JournalTitle{\apjl},
  723, L34}

\bibitem[{{D'Alessio} {et~al.}(2001){D'Alessio}, {Calvet}, \&
  {Hartmann}}]{Dalessio01}
{D'Alessio}, P., {Calvet}, N., \& {Hartmann}, L. 2001,
  \href{http://dx.doi.org/10.1086/320655}{\JournalTitle{\apj}, 553, 321}

\bibitem[{{D'Alessio} {et~al.}(1998){D'Alessio}, {Cant{\'o}}, {Calvet}, \&
  {Lizano}}]{Dalessio98}
{D'Alessio}, P., {Cant{\'o}}, J., {Calvet}, N., \& {Lizano}, S. 1998,
  \href{http://dx.doi.org/10.1086/305702}{\JournalTitle{\apj}, 500, 411}

\bibitem[{{Di Francesco} {et~al.}(2013){Di Francesco}, {Johnstone}, {Matthews},
  {Bartel}, {Bronfman}, {Casassus}, {Chitsazzadeh}, {Chou}, {Cunningham},
  {Duchene}, {Geisbuesch}, {Hales}, {Ho}, {Houde}, {Iono}, {Kemper}, {Kepley},
  {Koch}, {Kohno}, {Kothes}, {Lai}, {Lin}, {Liu}, {Mason}, {Maccarone},
  {Mizuno}, {Morata}, {Schieven}, {Scaife}, {Scott}, {Shang}, {Shimojo}, {Su},
  {Takakuwa}, {Wagg}, {Wootten}, \& {Yusef-Zadeh}}]{DiFrancesco13}
{Di Francesco}, J., {Johnstone}, D., {Matthews}, B.~C., {et~al.} 2013,
  \JournalTitle{ArXiv e-prints},
  \href{http://arxiv.org/abs/1310.1604}{{\sffamily arXiv:1310.1604
  [astro-ph.IM]}}

\bibitem[{{Draine}(2006)}]{Draine06}
{Draine}, B.~T. 2006,
  \href{http://dx.doi.org/10.1086/498130}{\JournalTitle{\apj}, 636, 1114}

\bibitem[{{Dunham} {et~al.}(2014){Dunham}, {Stutz}, {Allen}, {Evans},
  {Fischer}, {Megeath}, {Myers}, {Offner}, {Poteet}, {Tobin}, \&
  {Vorobyov}}]{Dunham14}
{Dunham}, M.~M., {Stutz}, A.~M., {Allen}, L.~E., {et~al.} 2014,
  \href{http://dx.doi.org/10.2458/azu_uapress_9780816531240-ch009}{\JournalTitle{Protostars
  and Planets VI}, 195}

\bibitem[{{Evans} {et~al.}(2017){Evans}, {Ilee}, {Hartquist}, {Caselli}, {Sz{\H
  u}cs}, {Purser}, {Boley}, {Durisen}, \& {Rawlings}}]{Evans17}
{Evans}, M.~G., {Ilee}, J.~D., {Hartquist}, T.~W., {et~al.} 2017,
  \href{http://dx.doi.org/10.1093/mnras/stx1365}{\JournalTitle{\mnras}, 470,
  1828}

\bibitem[{{Gerin} {et~al.}(2017){Gerin}, {Pety}, {Commer{\c c}on}, {Fuente},
  {Cernicharo}, {Marcelino}, {Ciardi}, {Lis}, {Roueff}, {Wootten}, \&
  {Chapillon}}]{Gerin17}
{Gerin}, M., {Pety}, J., {Commer{\c c}on}, B., {et~al.} 2017,
  \href{http://dx.doi.org/10.1051/0004-6361/201630187}{\JournalTitle{\aap},
  606, A35}

\bibitem[{{Girart} {et~al.}(2018){Girart}, {Fern{\'a}ndez-L{\'o}pez}, {Li},
  {Yang}, {Estalella}, {Anglada}, {{\'A}{\~n}ez-L{\'o}pez}, {Busquet},
  {Carrasco-Gonz{\'a}lez}, {Curiel}, {Galvan-Madrid}, {G{\'o}mez}, {de
  Gregorio-Monsalvo}, {Jim{\'e}nez-Serra}, {Krasnopolsky}, {Mart{\'{\i}}},
  {Osorio}, {Padovani}, {Rao}, {Rodr{\'{\i}}guez}, \& {Torrelles}}]{Girart18}
{Girart}, J.~M., {Fern{\'a}ndez-L{\'o}pez}, M., {Li}, Z.-Y., {et~al.} 2018,
  \href{http://dx.doi.org/10.3847/2041-8213/aab76b}{\JournalTitle{\apjl}, 856,
  L27}

\bibitem[{{Gr{\"a}fe} {et~al.}(2013){Gr{\"a}fe}, {Wolf}, {Guilloteau},
  {Dutrey}, {Stapelfeldt}, {Pontoppidan}, \& {Sauter}}]{Graefe13}
{Gr{\"a}fe}, C., {Wolf}, S., {Guilloteau}, S., {et~al.} 2013,
  \href{http://dx.doi.org/10.1051/0004-6361/201220720}{\JournalTitle{\aap},
  553, A69}

\bibitem[{{Guilloteau} {et~al.}(2016){Guilloteau}, {Pi{\'e}tu}, {Chapillon},
  {Di Folco}, {Dutrey}, {Henning}, {Semenov}, {Birnstiel}, \&
  {Grosso}}]{Guilloteau16}
{Guilloteau}, S., {Pi{\'e}tu}, V., {Chapillon}, E., {et~al.} 2016,
  \href{http://dx.doi.org/10.1051/0004-6361/201527620}{\JournalTitle{\aap},
  586, L1}

\bibitem[{{Hirano} \& {Liu}(2014)}]{Hirano14}
{Hirano}, N., \& {Liu}, F.-c. 2014,
  \href{http://dx.doi.org/10.1088/0004-637X/789/1/50}{\JournalTitle{\apj}, 789,
  50}

\bibitem[{{Ho} {et~al.}(2004){Ho}, {Moran}, \& {Lo}}]{Ho04}
{Ho}, P.~T.~P., {Moran}, J.~M., \& {Lo}, K.~Y. 2004,
  \href{http://dx.doi.org/10.1086/423245}{\JournalTitle{\apjl}, 616, L1}

\bibitem[{{Hosokawa} {et~al.}(2010){Hosokawa}, {Yorke}, \&
  {Omukai}}]{HosokawaOmukai10}
{Hosokawa}, T., {Yorke}, H.~W., \& {Omukai}, K. 2010,
  \href{http://dx.doi.org/10.1088/0004-637X/721/1/478}{\JournalTitle{\apj},
  721, 478}

\bibitem[{{Hull} {et~al.}(2018){Hull}, {Yang}, {Li}, {Kataoka}, {Stephens},
  {Andrews}, {Bai}, {Cleeves}, {Hughes}, {Looney}, {P{\'e}rez}, \&
  {Wilner}}]{Hull18}
{Hull}, C.~L.~H., {Yang}, H., {Li}, Z.-Y., {et~al.} 2018,
  \href{http://dx.doi.org/10.3847/1538-4357/aabfeb}{\JournalTitle{\apj}, 860,
  82}

\bibitem[{{Ilee} {et~al.}(2017){Ilee}, {Forgan}, {Evans}, {Hall}, {Booth},
  {Clarke}, {Rice}, {Boley}, {Caselli}, {Hartquist}, \& {Rawlings}}]{Ilee17}
{Ilee}, J.~D., {Forgan}, D.~H., {Evans}, M.~G., {et~al.} 2017,
  \href{http://dx.doi.org/10.1093/mnras/stx1966}{\JournalTitle{\mnras}, 472,
  189}

\bibitem[{{Isella} {et~al.}(2015){Isella}, {Hull}, {Moullet},
  {Galv{\'a}n-Madrid}, {Johnstone}, {Ricci}, {Tobin}, {Testi}, {Beltran},
  {Lazio}, {Siemion}, {Liu}, {Du}, {{\"O}berg}, {Bergin}, {Caselli}, {Bourke},
  {Carilli}, {Perez}, {Butler}, {de Pater}, {Qi}, {Hofstadter}, {Moreno},
  {Alexander}, {Williams}, {Goldsmith}, {Wyatt}, {Loinard}, {Di Francesco},
  {Wilner}, {Schilke}, {Ginsburg}, {S{\'a}nchez-Monge}, {Zhang}, \&
  {Beuther}}]{Isella15}
{Isella}, A., {Hull}, C.~L.~H., {Moullet}, A., {et~al.} 2015,
  \JournalTitle{ArXiv e-prints},
  \href{http://arxiv.org/abs/1510.06444}{{\sffamily arXiv:1510.06444
  [astro-ph.SR]}}

\bibitem[{{Izquierdo} {et~al.}(2018){Izquierdo}, {Galv{\'a}n-Madrid}, {Maud},
  {Hoare}, {Johnston}, {Keto}, {Zhang}, \& {de Wit}}]{Izquierdo18}
{Izquierdo}, A.~F., {Galv{\'a}n-Madrid}, R., {Maud}, L.~T., {et~al.} 2018,
  \href{http://dx.doi.org/10.1093/mnras/sty1096}{\JournalTitle{\mnras}, 478,
  2505}

\bibitem[{{J{\o}rgensen} {et~al.}(2009){J{\o}rgensen}, {van Dishoeck},
  {Visser}, {Bourke}, {Wilner}, {Lommen}, {Hogerheijde}, \&
  {Myers}}]{Jorgensen09}
{J{\o}rgensen}, J.~K., {van Dishoeck}, E.~F., {Visser}, R., {et~al.} 2009,
  \href{http://dx.doi.org/10.1051/0004-6361/200912325}{\JournalTitle{\aap},
  507, 861}

\bibitem[{{Kataoka} {et~al.}(2016){Kataoka}, {Muto}, {Momose}, {Tsukagoshi}, \&
  {Dullemond}}]{Kataoka16}
{Kataoka}, A., {Muto}, T., {Momose}, M., {Tsukagoshi}, T., \& {Dullemond},
  C.~P. 2016,
  \href{http://dx.doi.org/10.3847/0004-637X/820/1/54}{\JournalTitle{\apj}, 820,
  54}

\bibitem[{{Kataoka} {et~al.}(2015){Kataoka}, {Muto}, {Momose}, {Tsukagoshi},
  {Fukagawa}, {Shibai}, {Hanawa}, {Murakawa}, \& {Dullemond}}]{Kataoka15}
{Kataoka}, A., {Muto}, T., {Momose}, M., {et~al.} 2015,
  \href{http://dx.doi.org/10.1088/0004-637X/809/1/78}{\JournalTitle{\apj}, 809,
  78}

\bibitem[{{Keto} \& {Zhang}(2010)}]{Keto10}
{Keto}, E., \& {Zhang}, Q. 2010,
  \href{http://dx.doi.org/10.1111/j.1365-2966.2010.16672.x}{\JournalTitle{\mnras},
  406, 102}

\bibitem[{{Kroupa}(2001)}]{Kroupa01}
{Kroupa}, P. 2001,
  \href{http://dx.doi.org/10.1046/j.1365-8711.2001.04022.x}{\JournalTitle{\mnras},
  322, 231}

\bibitem[{{Lee}(2010)}]{Lee10}
{Lee}, C.-F. 2010,
  \href{http://dx.doi.org/10.1088/0004-637X/725/1/712}{\JournalTitle{\apj},
  725, 712}

\bibitem[{{Lee} {et~al.}(2017{\natexlab{a}}){Lee}, {Ho}, {Li}, {Hirano},
  {Zhang}, \& {Shang}}]{Lee17a}
{Lee}, C.-F., {Ho}, P.~T.~P., {Li}, Z.-Y., {et~al.} 2017{\natexlab{a}},
  \href{http://dx.doi.org/10.1038/s41550-017-0152}{\JournalTitle{Nature
  Astronomy}, 1, 0152}

\bibitem[{{Lee} {et~al.}(2018){Lee}, {Li}, {Ching}, {Lai}, \& {Yang}}]{Lee18}
{Lee}, C.-F., {Li}, Z.-Y., {Ching}, T.-C., {Lai}, S.-P., \& {Yang}, H. 2018,
  \href{http://dx.doi.org/10.3847/1538-4357/aaa769}{\JournalTitle{\apj}, 854,
  56}

\bibitem[{{Lee} {et~al.}(2017{\natexlab{b}}){Lee}, {Li}, {Ho}, {Hirano},
  {Zhang}, \& {Shang}}]{Lee17b}
{Lee}, C.-F., {Li}, Z.-Y., {Ho}, P.~T.~P., {et~al.} 2017{\natexlab{b}},
  \href{http://dx.doi.org/10.1126/sciadv.1602935}{\JournalTitle{Science
  Advances}, 3, e1602935}

\bibitem[{{Li} {et~al.}(2017){Li}, {Liu}, {Hasegawa}, \& {Hirano}}]{LiLiu17}
{Li}, J.~I., {Liu}, H.~B., {Hasegawa}, Y., \& {Hirano}, N. 2017,
  \href{http://dx.doi.org/10.3847/1538-4357/aa6f04}{\JournalTitle{\apj}, 840,
  72}

\bibitem[{{Li} {et~al.}(2014){Li}, {Banerjee}, {Pudritz}, {J{\o}rgensen},
  {Shang}, {Krasnopolsky}, \& {Maury}}]{Li14}
{Li}, Z.-Y., {Banerjee}, R., {Pudritz}, R.~E., {et~al.} 2014,
  \href{http://dx.doi.org/10.2458/azu_uapress_9780816531240-ch008}{\JournalTitle{Protostars
  and Planets VI}, 173}

\bibitem[{{Liu} {et~al.}(2017){Liu}, {Vorobyov}, {Dong}, {Dunham}, {Takami},
  {Galv{\'a}n-Madrid}, {Hashimoto}, {K{\'o}sp{\'a}l}, {Henning}, {Tamura},
  {Rodr{\'{\i}}guez}, {Hirano}, {Hasegawa}, {Fukagawa}, {Carrasco-Gonzalez}, \&
  {Tazzari}}]{Liu17}
{Liu}, H.~B., {Vorobyov}, E.~I., {Dong}, R., {et~al.} 2017,
  \href{http://dx.doi.org/10.1051/0004-6361/201630263}{\JournalTitle{\aap},
  602, A19}

\bibitem[{{Liu} {et~al.}(2018){Liu}, {Dunham}, {Pascucci}, {Bourke}, {Hirano},
  {Longmore}, {Andrews}, {Carrasco-Gonz{\'a}lez}, {Forbrich},
  {Galv{\'a}n-Madrid}, {Girart}, {Green}, {Ju{\'a}rez}, {K{\'o}sp{\'a}l},
  {Manara}, {Palau}, {Takami}, {Testi}, \& {Vorobyov}}]{Liu18}
{Liu}, H.~B., {Dunham}, M.~M., {Pascucci}, I., {et~al.} 2018,
  \href{http://dx.doi.org/10.1051/0004-6361/201731951}{\JournalTitle{\aap},
  612, A54}

\bibitem[{{Lizano} \& {Galli}(2015)}]{LizanoGalli15}
{Lizano}, S., \& {Galli}, D. 2015,
  \href{http://dx.doi.org/10.1007/978-3-662-44625-6_16}{in Astrophysics and
  Space Science Library, Vol. 407, Magnetic Fields in Diffuse Media, ed.
  A.~{Lazarian}, E.~M. {de Gouveia Dal Pino}, \& C.~{Melioli}}, 459

\bibitem[{{Machida} {et~al.}(2014){Machida}, {Inutsuka}, \&
  {Matsumoto}}]{Machida14}
{Machida}, M.~N., {Inutsuka}, S.-i., \& {Matsumoto}, T. 2014,
  \href{http://dx.doi.org/10.1093/mnras/stt2343}{\JournalTitle{\mnras}, 438,
  2278}

\bibitem[{{McClure} {et~al.}(2016){McClure}, {Bergin}, {Cleeves}, {van
  Dishoeck}, {Blake}, {Evans}, {Green}, {Henning}, {{\"O}berg}, {Pontoppidan},
  \& {Salyk}}]{McClure16}
{McClure}, M.~K., {Bergin}, E.~A., {Cleeves}, L.~I., {et~al.} 2016,
  \href{http://dx.doi.org/10.3847/0004-637X/831/2/167}{\JournalTitle{\apj},
  831, 167}

\bibitem[{{Miller} \& {Scalo}(1979)}]{Miller1979}
{Miller}, G.~E., \& {Scalo}, J.~M. 1979,
  \href{http://dx.doi.org/10.1086/190629}{\JournalTitle{\apjs}, 41, 513}

\bibitem[{{Miotello} {et~al.}(2014){Miotello}, {Testi}, {Lodato}, {Ricci},
  {Rosotti}, {Brooks}, {Maury}, \& {Natta}}]{Miotello14}
{Miotello}, A., {Testi}, L., {Lodato}, G., {et~al.} 2014,
  \href{http://dx.doi.org/10.1051/0004-6361/201322945}{\JournalTitle{\aap},
  567, A32}

\bibitem[{{Miotello} {et~al.}(2017){Miotello}, {van Dishoeck}, {Williams},
  {Ansdell}, {Guidi}, {Hogerheijde}, {Manara}, {Tazzari}, {Testi}, {van der
  Marel}, \& {van Terwisga}}]{Miotello17}
{Miotello}, A., {van Dishoeck}, E.~F., {Williams}, J.~P., {et~al.} 2017,
  \href{http://dx.doi.org/10.1051/0004-6361/201629556}{\JournalTitle{\aap},
  599, A113}

\bibitem[{{Murillo} {et~al.}(2013){Murillo}, {Lai}, {Bruderer}, {Harsono}, \&
  {van Dishoeck}}]{Murillo13}
{Murillo}, N.~M., {Lai}, S.-P., {Bruderer}, S., {Harsono}, D., \& {van
  Dishoeck}, E.~F. 2013,
  \href{http://dx.doi.org/10.1051/0004-6361/201322537}{\JournalTitle{\aap},
  560, A103}

\bibitem[{{Ormel} {et~al.}(2009){Ormel}, {Paszun}, {Dominik}, \&
  {Tielens}}]{Ormel09}
{Ormel}, C.~W., {Paszun}, D., {Dominik}, C., \& {Tielens}, A.~G.~G.~M. 2009,
  \href{http://dx.doi.org/10.1051/0004-6361/200811158}{\JournalTitle{\aap},
  502, 845}

\bibitem[{{Pascucci} {et~al.}(2016){Pascucci}, {Testi}, {Herczeg}, {Long},
  {Manara}, {Hendler}, {Mulders}, {Krijt}, {Ciesla}, {Henning}, {Mohanty},
  {Drabek-Maunder}, {Apai}, {Sz{\H u}cs}, {Sacco}, \& {Olofsson}}]{Pascucci16}
{Pascucci}, I., {Testi}, L., {Herczeg}, G.~J., {et~al.} 2016,
  \href{http://dx.doi.org/10.3847/0004-637X/831/2/125}{\JournalTitle{\apj},
  831, 125}

\bibitem[{{Pringle}(1981)}]{Pringle81}
{Pringle}, J.~E. 1981,
  \href{http://dx.doi.org/10.1146/annurev.aa.19.090181.001033}{\JournalTitle{\araa},
  19, 137}

\bibitem[{{Robitaille} \& {Bressert}(2012)}]{AplPy}
{Robitaille}, T., \& {Bressert}, E. 2012, {APLpy: Astronomical Plotting Library
  in Python}, Astrophysics Source Code Library,
  \href{http://arxiv.org/abs/1208.017}{{\sffamily ascl:1208.017}}

\bibitem[{{Rodr{\'{\i}}guez} {et~al.}(1998){Rodr{\'{\i}}guez}, {D'Alessio},
  {Wilner}, {Ho}, {Torrelles}, {Curiel}, {G{\'o}mez}, {Lizano}, {Pedlar},
  {Cant{\'o}}, \& {Raga}}]{Rodriguez98}
{Rodr{\'{\i}}guez}, L.~F., {D'Alessio}, P., {Wilner}, D.~J., {et~al.} 1998,
  \href{http://dx.doi.org/10.1038/26421}{\JournalTitle{\nat}, 395, 355}

\bibitem[{{Sakai} {et~al.}(2014){Sakai}, {Sakai}, {Hirota}, {Watanabe},
  {Ceccarelli}, {Kahane}, {Bottinelli}, {Caux}, {Demyk}, {Vastel}, {Coutens},
  {Taquet}, {Ohashi}, {Takakuwa}, {Yen}, {Aikawa}, \& {Yamamoto}}]{Sakai14}
{Sakai}, N., {Sakai}, T., {Hirota}, T., {et~al.} 2014,
  \href{http://dx.doi.org/10.1038/nature13000}{\JournalTitle{\nat}, 507, 78}

\bibitem[{{Salpeter}(1955)}]{Salpeter55}
{Salpeter}, E.~E. 1955,
  \href{http://dx.doi.org/10.1086/145971}{\JournalTitle{\apj}, 121, 161}

\bibitem[{{Seifried} {et~al.}(2015){Seifried}, {Banerjee}, {Pudritz}, \&
  {Klessen}}]{Seifried15}
{Seifried}, D., {Banerjee}, R., {Pudritz}, R.~E., \& {Klessen}, R.~S. 2015,
  \href{http://dx.doi.org/10.1093/mnras/stu2282}{\JournalTitle{\mnras}, 446,
  2776}

\bibitem[{{Sheehan} \& {Eisner}(2017)}]{Sheehan2017}
{Sheehan}, P.~D., \& {Eisner}, J.~A. 2017,
  \href{http://dx.doi.org/10.3847/1538-4357/aa9990}{\JournalTitle{\apj}, 851,
  45}

\bibitem[{{Tazzari} {et~al.}(2016){Tazzari}, {Testi}, {Ercolano}, {Natta},
  {Isella}, {Chandler}, {P{\'e}rez}, {Andrews}, {Wilner}, {Ricci}, {Henning},
  {Linz}, {Kwon}, {Corder}, {Dullemond}, {Carpenter}, {Sargent}, {Mundy},
  {Storm}, {Calvet}, {Greaves}, {Lazio}, \& {Deller}}]{Tazzari16}
{Tazzari}, M., {Testi}, L., {Ercolano}, B., {et~al.} 2016,
  \href{http://dx.doi.org/10.1051/0004-6361/201527423}{\JournalTitle{\aap},
  588, A53}

\bibitem[{{Testi} {et~al.}(2014){Testi}, {Birnstiel}, {Ricci}, {Andrews},
  {Blum}, {Carpenter}, {Dominik}, {Isella}, {Natta}, {Williams}, \&
  {Wilner}}]{Testi14}
{Testi}, L., {Birnstiel}, T., {Ricci}, L., {et~al.} 2014,
  \href{http://dx.doi.org/10.2458/azu_uapress_9780816531240-ch015}{\JournalTitle{Protostars
  and Planets VI}, 339}

\bibitem[{{Tobin} {et~al.}(2013){Tobin}, {Hartmann}, {Chiang}, {Wilner},
  {Looney}, {Loinard}, {Calvet}, \& {D'Alessio}}]{Tobin13}
{Tobin}, J.~J., {Hartmann}, L., {Chiang}, H.-F., {et~al.} 2013,
  \href{http://dx.doi.org/10.1088/0004-637X/771/1/48}{\JournalTitle{\apj}, 771,
  48}

\bibitem[{{Tobin} {et~al.}(2016){Tobin}, {Kratter}, {Persson}, {Looney},
  {Dunham}, {Segura-Cox}, {Li}, {Chandler}, {Sadavoy}, {Harris}, {Melis}, \&
  {P{\'e}rez}}]{Tobin16}
{Tobin}, J.~J., {Kratter}, K.~M., {Persson}, M.~V., {et~al.} 2016,
  \href{http://dx.doi.org/10.1038/nature20094}{\JournalTitle{\nat}, 538, 483}

\bibitem[{{Van Der Walt} {et~al.}(2011){Van Der Walt}, {Colbert}, \&
  {Varoquaux}}]{Numpy}
{Van Der Walt}, S., {Colbert}, S.~C., \& {Varoquaux}, G. 2011,
  \JournalTitle{ArXiv e-prints},
  \href{http://arxiv.org/abs/1102.1523}{{\sffamily arXiv:1102.1523 [cs.MS]}}

\bibitem[{{Vorobyov} {et~al.}(2018){Vorobyov}, {Akimkin}, {Stoyanovskaya},
  {Pavlyuchenkov}, \& {Liu}}]{Vorobyov18}
{Vorobyov}, E.~I., {Akimkin}, V., {Stoyanovskaya}, O., {Pavlyuchenkov}, Y., \&
  {Liu}, H.~B. 2018,
  \href{http://dx.doi.org/10.1051/0004-6361/201731690}{\JournalTitle{\aap},
  614, A98}

\bibitem[{{Vorobyov} {et~al.}(2015){Vorobyov}, {Lin}, \& {Guedel}}]{Vorobyov15}
{Vorobyov}, E.~I., {Lin}, D.~N.~C., \& {Guedel}, M. 2015,
  \href{http://dx.doi.org/10.1051/0004-6361/201424583}{\JournalTitle{\aap},
  573, A5}

\bibitem[{{Whitney} {et~al.}(2003){Whitney}, {Wood}, {Bjorkman}, \&
  {Wolff}}]{Whitney03}
{Whitney}, B.~A., {Wood}, K., {Bjorkman}, J.~E., \& {Wolff}, M.~J. 2003,
  \href{http://dx.doi.org/10.1086/375415}{\JournalTitle{\apj}, 591, 1049}

\bibitem[{{Wong} {et~al.}(2016){Wong}, {Hirashita}, \& {Li}}]{Wong16}
{Wong}, Y.~H.~V., {Hirashita}, H., \& {Li}, Z.-Y. 2016,
  \href{http://dx.doi.org/10.1093/pasj/psw066}{\JournalTitle{\pasj}, 68, 67}

\bibitem[{{Yang} {et~al.}(2016){Yang}, {Li}, {Looney}, \& {Stephens}}]{Yang16}
{Yang}, H., {Li}, Z.-Y., {Looney}, L., \& {Stephens}, I. 2016,
  \href{http://dx.doi.org/10.1093/mnras/stv2633}{\JournalTitle{\mnras}, 456,
  2794}

\bibitem[{{Zhao} {et~al.}(2018){Zhao}, {Caselli}, {Li}, \&
  {Krasnopolsky}}]{Zhao18}
{Zhao}, B., {Caselli}, P., {Li}, Z.-Y., \& {Krasnopolsky}, R. 2018,
  \href{http://dx.doi.org/10.1093/mnras/stx2617}{\JournalTitle{\mnras}, 473,
  4868}

\bibitem[{{Zhao} {et~al.}(2016){Zhao}, {Caselli}, {Li}, {Krasnopolsky},
  {Shang}, \& {Nakamura}}]{Zhao16}
{Zhao}, B., {Caselli}, P., {Li}, Z.-Y., {et~al.} 2016,
  \href{http://dx.doi.org/10.1093/mnras/stw1124}{\JournalTitle{\mnras}, 460,
  2050}

\bibitem[{{Zhu} {et~al.}(2012){Zhu}, {Hartmann}, {Nelson}, \& {Gammie}}]{Zhu12}
{Zhu}, Z., {Hartmann}, L., {Nelson}, R.~P., \& {Gammie}, C.~F. 2012,
  \href{http://dx.doi.org/10.1088/0004-637X/746/1/110}{\JournalTitle{\apj},
  746, 110}

\end{thebibliography}

\appendix

\section{Fiducial Model Structure} 
\label{sec:fiducial_structure}

\begin{figure*}[!ht]
\gridline{\fig{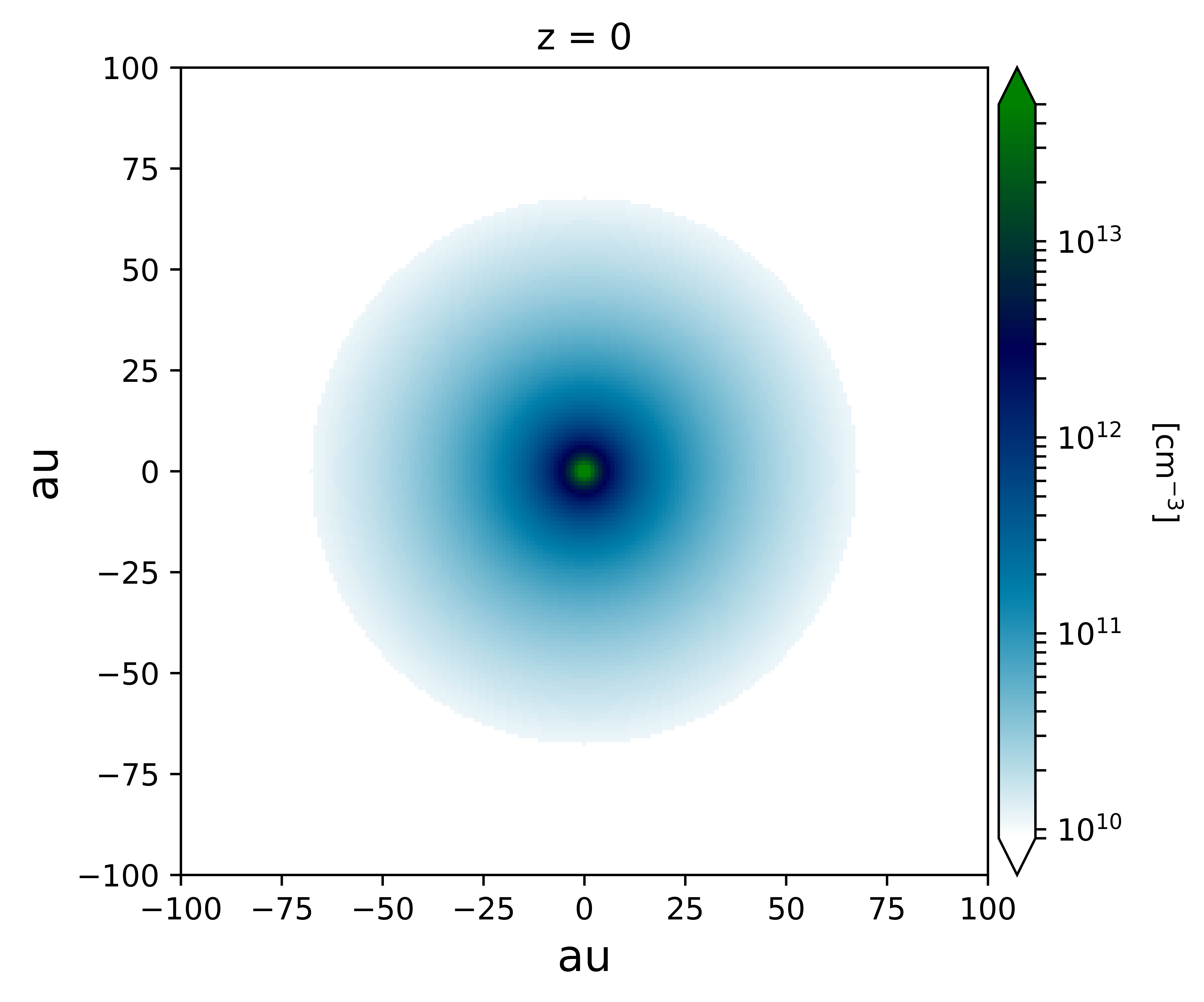}{0.5\textwidth}{{\it (a)}}
          \fig{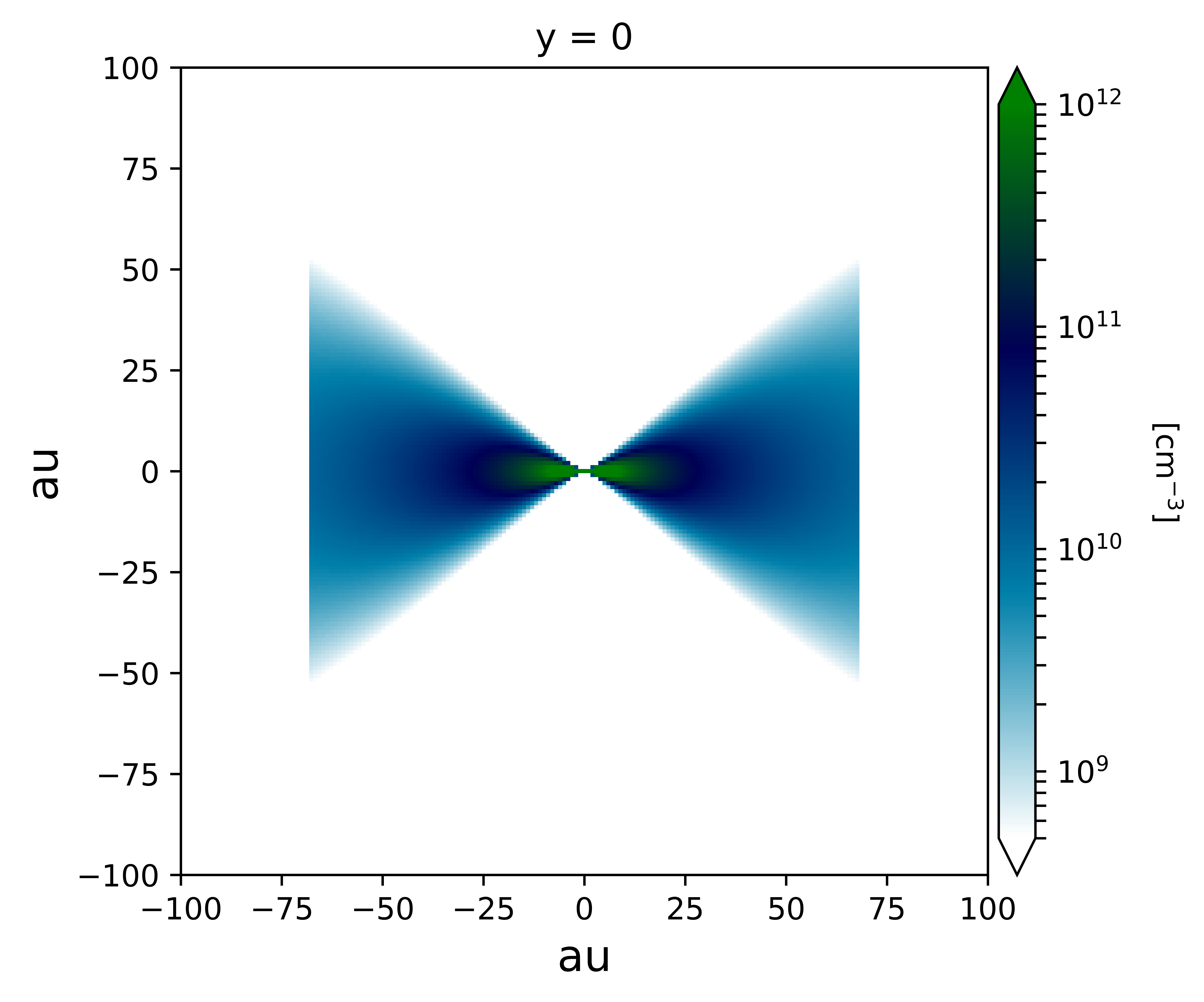}{0.5\textwidth}{{\it (b)}}
          }
\gridline{\fig{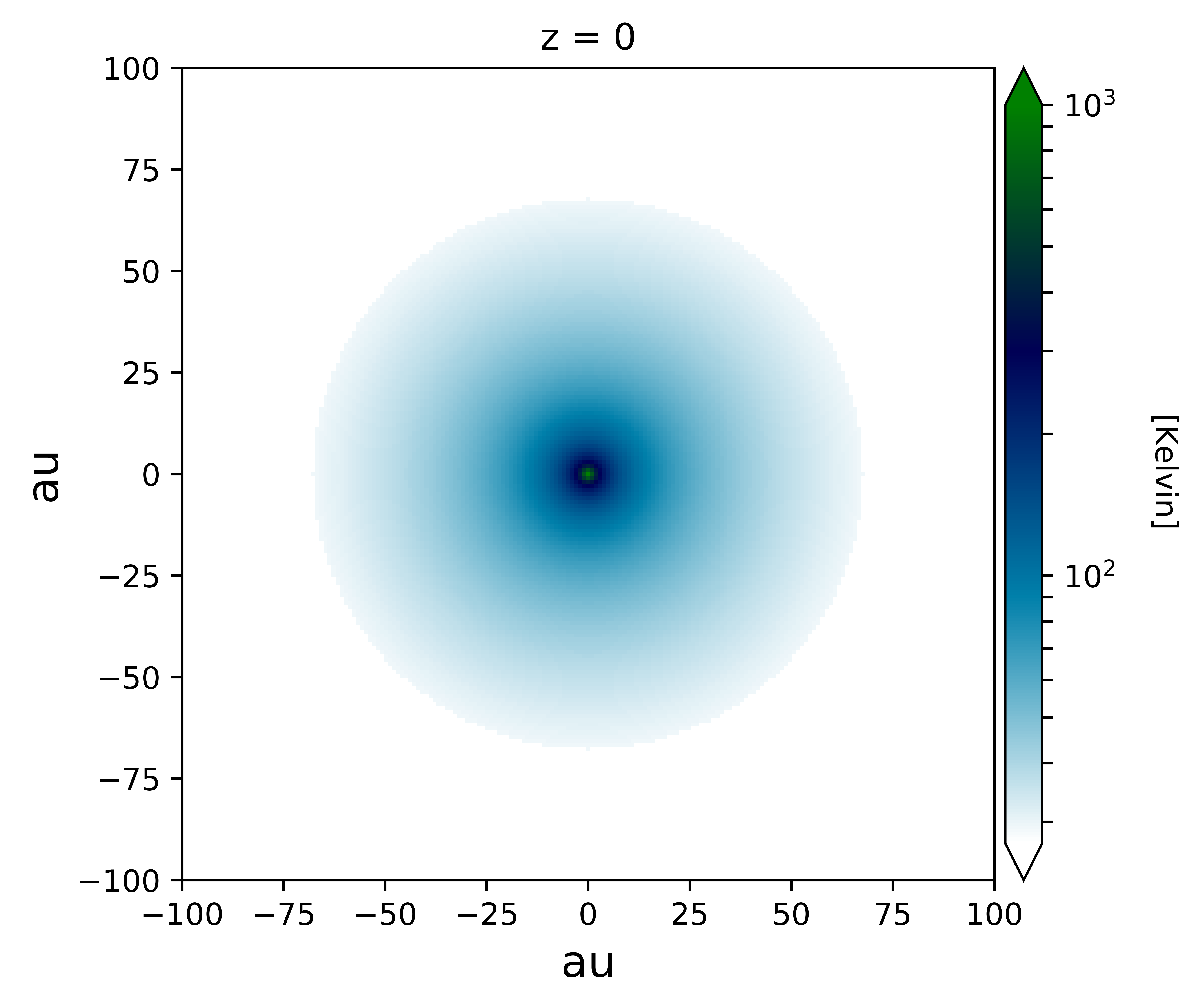}{0.5\textwidth}{{\it (c)}}
          \fig{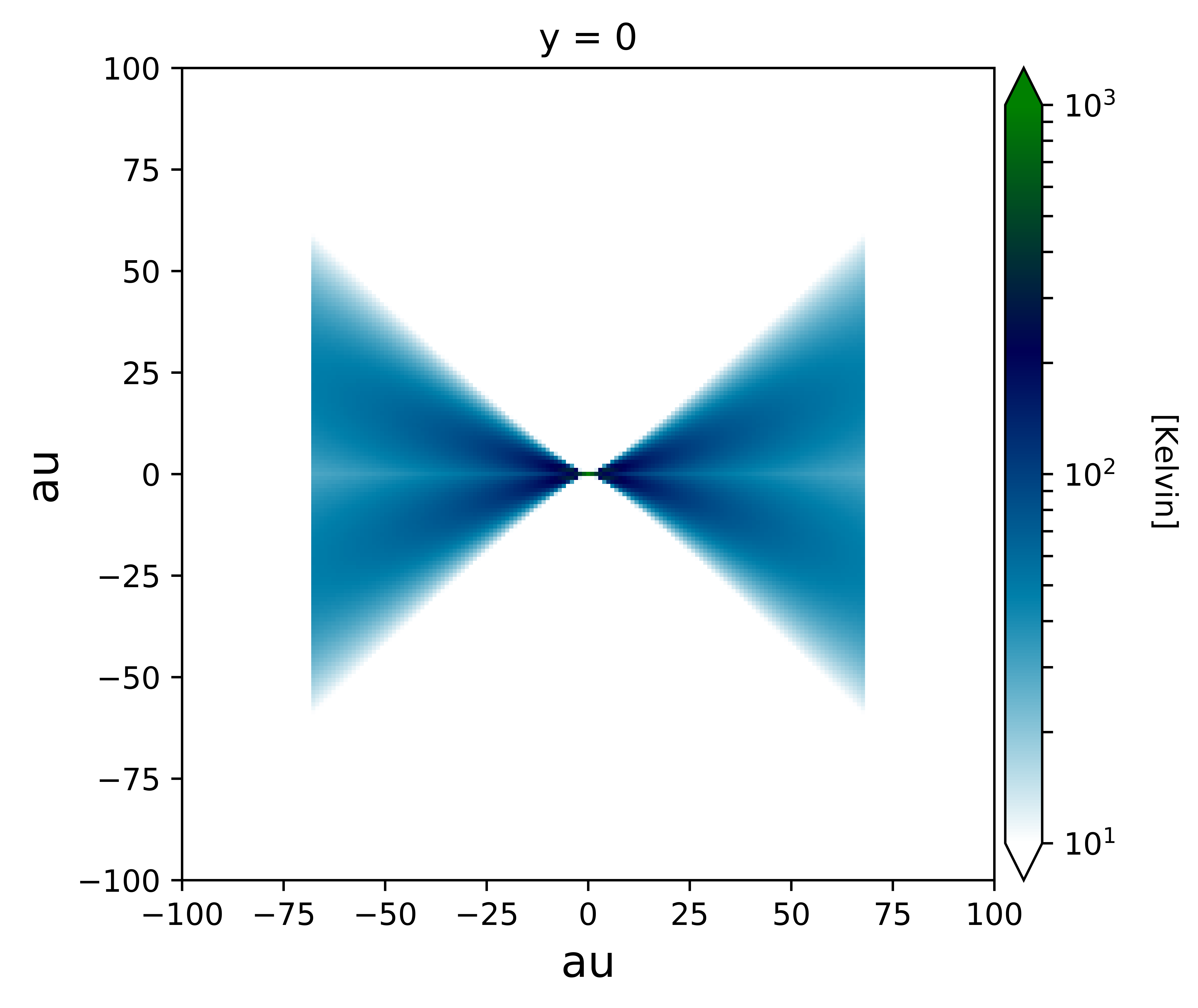}{0.5\textwidth}{{\it (d)}}
          }
\caption{{\it (a)} Midplane density. {\it (b)} Vertical density. {\it (c)} Midplane temperature. {\it (d)} Vertical temperature. \label{fig:model_cuts}}
\end{figure*}

\begin{figure*}[!ht]
\begin{center}
\includegraphics[width=0.51\textwidth]{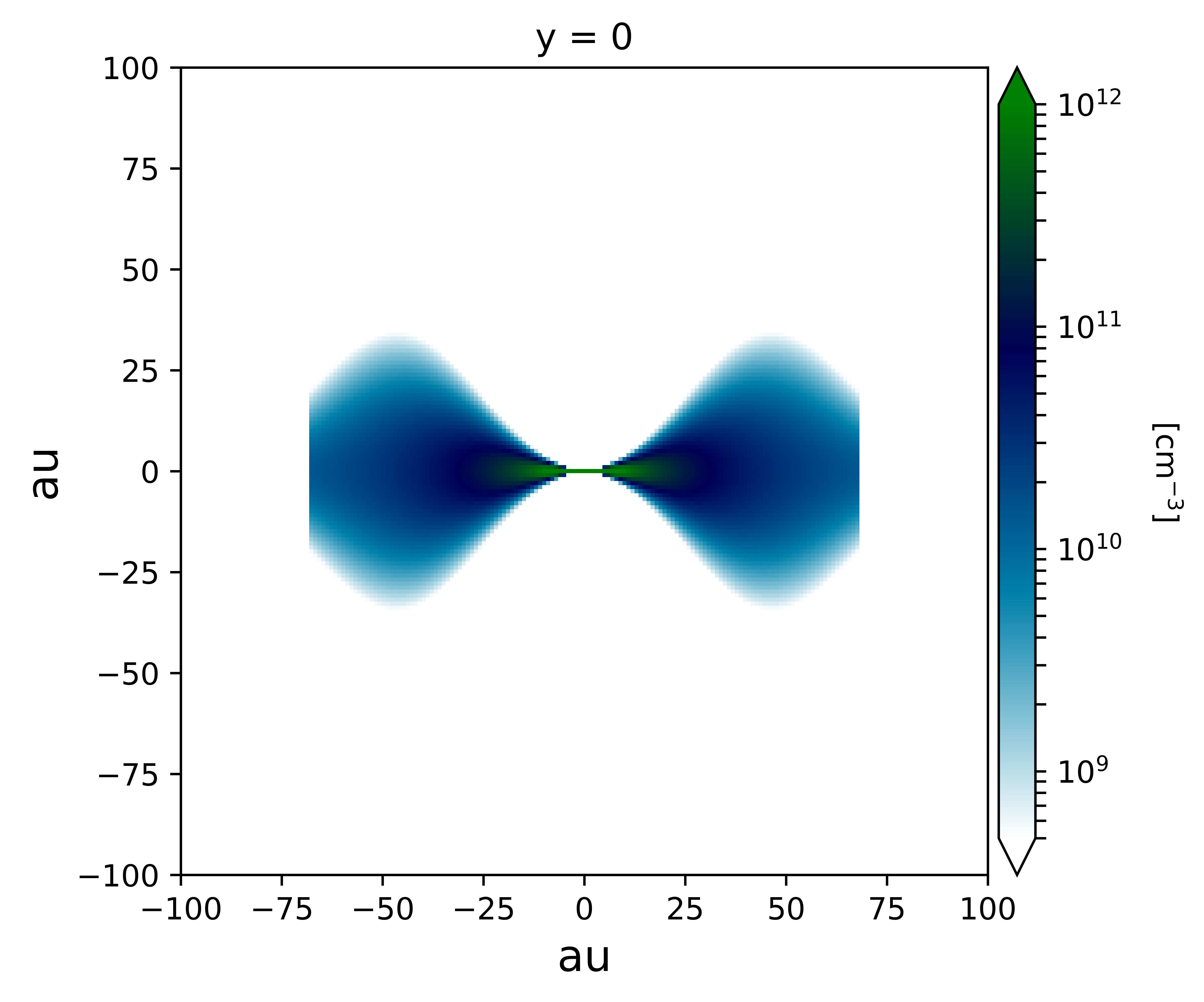}
\includegraphics[width=0.47\textwidth]{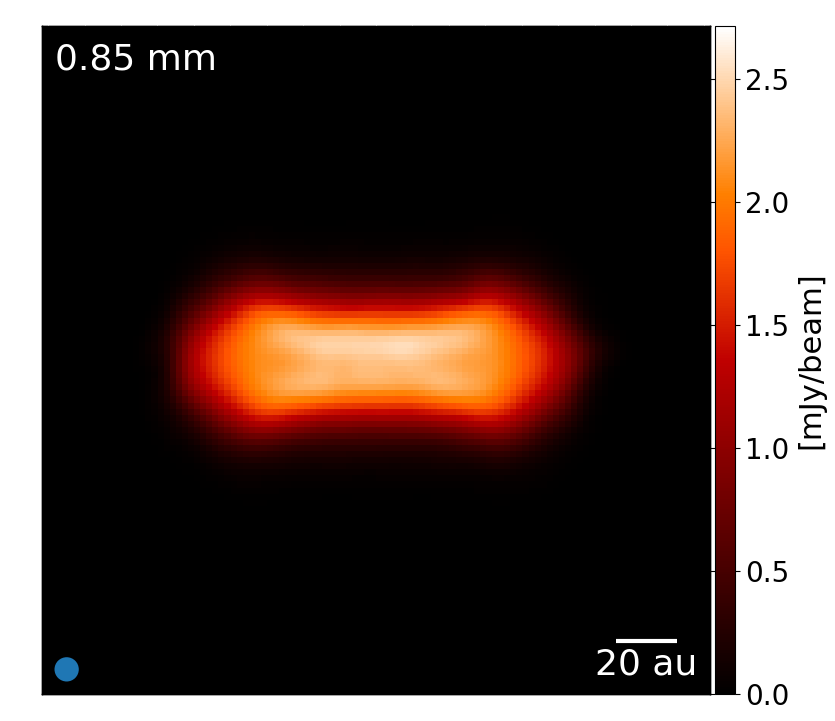}
\caption{
{\it Left:} midplane density of tapered model. {\it Right:} resulting 0.85 mm image as in Fig. \ref{fig:benchmark}. 
\label{fig:tapered_model}}
\end{center}
\end{figure*}

Figure \ref{fig:model_cuts} shows central cuts of density and temperature in the fiducial ($M_{\rm disk} = 0.15~M_\odot$) model midplane and perpendicular to it. Densities are in particles of H$_2$ cm$^{-3}$. Temperatures are in K. 
The density grids for all the other models scale linearly with the total model mass.
Figure \ref{fig:tapered_model} presents an example of a model with a an exponential tapering in the outer disk. 
We use the model library presented by \cite{Izquierdo18}, which includes a variety of model prescriptions other than the ones assumed in this paper.

\newpage

\newpage

\section{Compilation of Fluxes and Spectral Indices}
\label{sec:compilation}

\begin{figure*}[!ht]
\gridline{
\fig{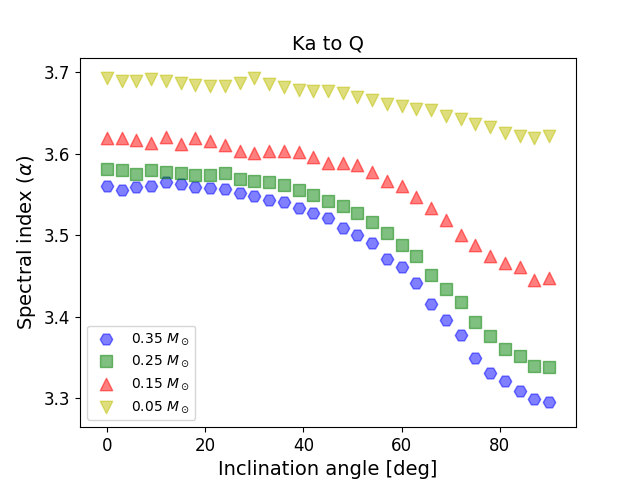}{0.34\textwidth}{}
\fig{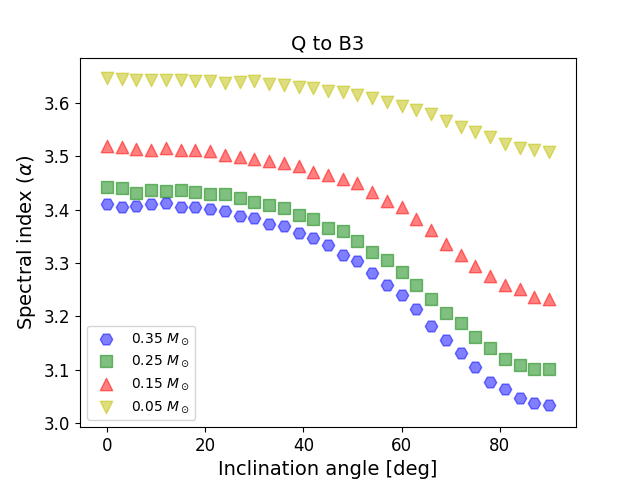}{0.34\textwidth}{}
\fig{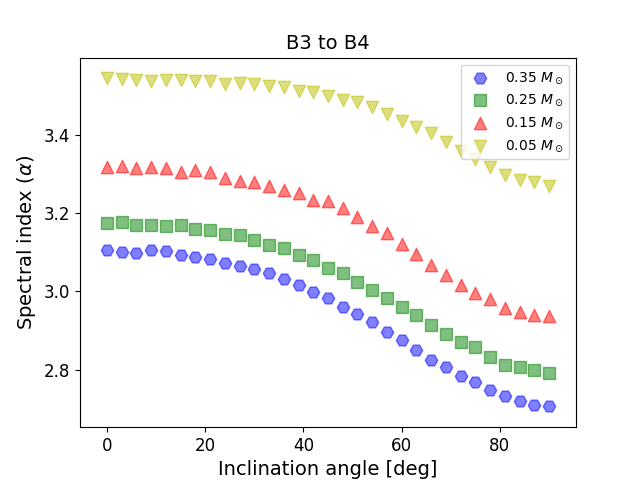}{0.34\textwidth}{}
          }
\gridline{
\fig{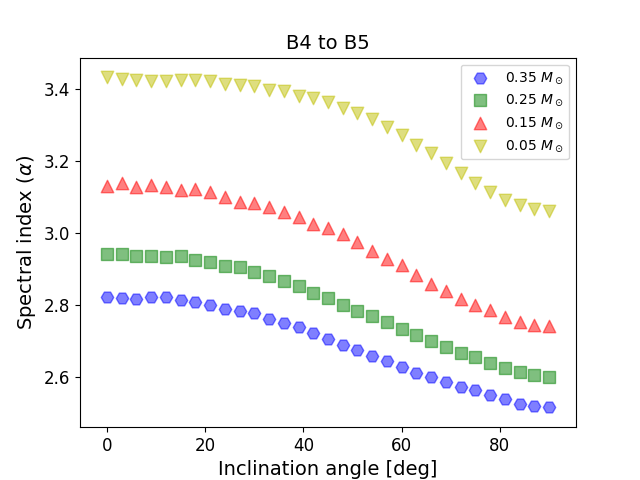}{0.34\textwidth}{}
\fig{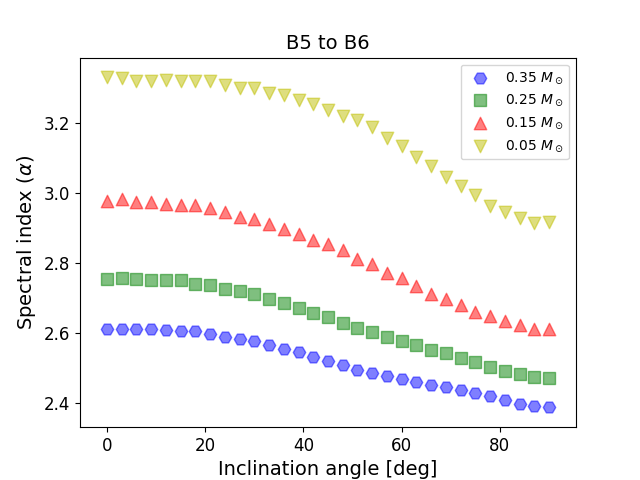}{0.34\textwidth}{}
\fig{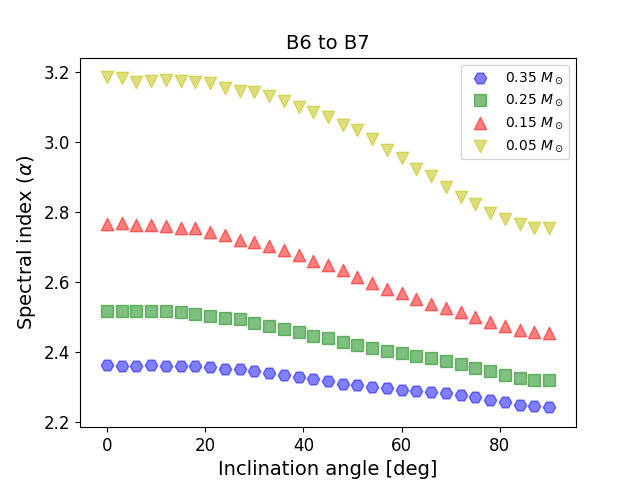}{0.34\textwidth}{}
}
\gridline{
\fig{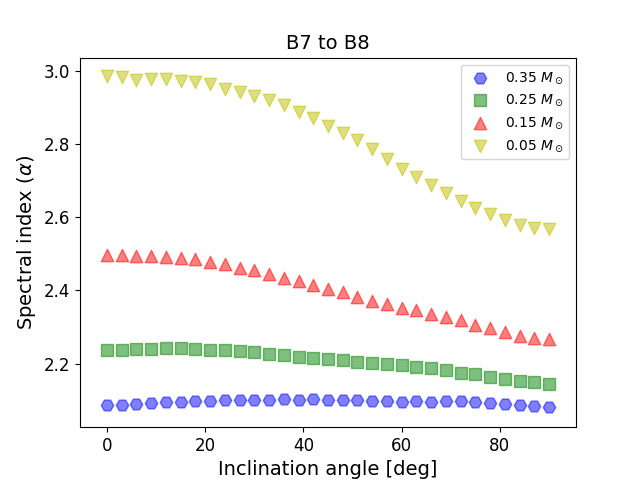}{0.34\textwidth}{}
          }
\caption{
Spectral indices for consecutive observing bands.  
\label{fig:compilation_alpha}}
\end{figure*}

Figure \ref{fig:compilation_alpha} shows the spectral indices for  additional frequency pairs than those presented in Fig. \ref{fig:representative_spix}. The file {\it fluxes\_hamburgers\_journal.txt} contains the background-subtracted model fluxes for the representative grid of models.

\end{document}